\documentclass[10pt,letterpaper]{article}
\pdfoutput=1

\usepackage{jcappub}
\usepackage{amsmath,amssymb,color}
\usepackage{enumerate,xstring}
\usepackage{jcappub}
\usepackage{amsmath}
\usepackage{amsfonts}
\usepackage{amssymb}
\usepackage{graphicx}
\usepackage{epsfig}
\usepackage{color}
\usepackage{multirow}
\usepackage{graphicx}
\usepackage{bigints}
\usepackage{todonotes,bm,etoolbox}
\usepackage{calligra}
\usepackage{hyperref}
\usepackage{tabularx}
\usepackage{booktabs}
\usepackage{subfigure}

\usepackage{tikz}

\def\ba#1\ea{\begin{align}#1\end{align}}
\def\bea{\begin{eqnarray}}
\def\eea{\end{eqnarray}}
\def\be{\begin{equation}}
\def\ee{\end{equation}}

\def\({\left(}
\def\){\right)}
\def\[{\left[}
\def\]{\right]}

\def\<{\left\langle}
\def\>{\right\rangle}

\def\comment#1{}

\def\eps{\epsilon}

\renewcommand{\v}[1]{\bm{#1}}

\def\vx{\v{x}}
\def\vk{\v{k}}
\def\vq{{\v{q}}}
\def\vp{\v{p}}

\def\vr{\v{r}}
\def\vtheta{\v{\theta}}
\def\vM{\v{M}}
\def\vD{\v{D}}

\newcommand{\perm}[1]{ \expandafter\ifstrempty\expandafter{#1} {\mbox{perm.}} {\mbox{$#1$ perm.}} }

\DeclareMathOperator{\Cov}{\rm \bf Cov}
\DeclareMathOperator{\cov}{\rm \bf Cov}

\newcommand{\fnl}{f_\textnormal{\textsc{nl}}}

\definecolor{RedWine}{rgb}{0.743,0,0}
\definecolor{RoyalBlue}{rgb}{0.25,.41,.88}
\definecolor{ForestGreen}{rgb}{.13,.54,.13}
\definecolor{Goldenrod}{rgb}{.85,.65,.13}

\newcommand{\bq}{\begin{eqnarray}}
\newcommand{\eq}{\end{eqnarray}}

\title{ \huge On the impact of galaxy bias uncertainties on primordial non-Gaussianity constraints}

\author[a]{Alexandre Barreira}

\affiliation[a]{Max-Planck-Institut f\"ur Astrophysik, Karl-Schwarzschild-Stra\ss e~1, 85748 Garching, Germany}

\emailAdd{barreira@mpa-garching.mpg.de}

\date{\today}

\abstract{We study the impact that uncertainties on assumed relations between galaxy bias parameters have on constraints of the local PNG $\fnl$ parameter. We focus on the relation between the linear density galaxy bias $b_1$ and local PNG bias $b_\phi$ in an idealized forecast setup with multitracer galaxy power spectrum and bispectrum data. We consider two parametrizations of galaxy bias: 1) one inspired by the universality relation where $b_\phi = 2\delta_c\left(b_1 - p\right)$ and $p$ is a free parameter; and 2) another in which the product of bias parameters and $\fnl$, like $\fnl b_\phi$, is directly fitted for. The constraints on the $\fnl-p$ plane are markedly bimodal, and both the central value and width of marginalized constraints on $\fnl$ depend sensitively on the priors on $p$.  Assuming fixed $p=1$ in the constraints with a fiducial value of $p=0.55$ can bias the inferred $\fnl$ by $0.5\sigma$ to $1\sigma$; priors $\Delta p \approx 0.5$ around this fiducial value are however sufficient in our setup to return unbiased constraints. In power spectrum analyses, parametrization 2, that makes no assumptions on $b_\phi$, can distinguish $\fnl \neq 0$ with the same significance as parametrization 1 assuming perfect knowledge of $b_\phi$ (the value of $\fnl$ is however left unknown). A drawback of parametrization 2 is that the addition of the bispectrum information is not as beneficial as in parametrization 1.  Our results motivate strongly the incorporation of mitigation strategies for bias uncertainties in PNG constraint analyses, as well as further theoretical studies on the relations between bias parameters to better inform those strategies.}

\begin{document}

\maketitle

\section{Introduction}
\label{sec:introduction}

One of the main goals of modern cosmology is to determine the statistical properties of the density fluctuations of the primordial universe and gain insights into the physics of the mechanisms that generated them during the epoch of inflation. Current observational constraints are compatible with the simplest models of inflation that involve a single scalar field slowly rolling down its potential. A key prediction of these models is that the resulting fluctuations should be Gaussian distributed \cite{maldacena:2003, 2004JCAP...10..006C, 2011JCAP...11..038C, Tanaka:2011aj, baldauf/etal:2011, conformalfermi, CFCpaper2, 2015JCAP...10..024D}, and consequently, studies of primordial non-Gaussianity (PNG) have become a major focus in theoretical and observational cosmology given its power to discriminate between single-field models and more elaborate models involving multiple fields (see Ref.~\cite{2014arXiv1412.4671A} for an overview). A popular characterization of PNG is that of the so-called {\it local} type, in which the primordial gravitational (Bardeen) potential $\phi(\vx)$ is expanded as \cite{2001PhRvD..63f3002K}
\bq\label{eq:fnl}
\phi(\vx) = \phi_{\rm G}(\vx) + \fnl\left[\phi_{\rm G}(\vx)^2 - \left<\phi_{\rm G}(\vx)^2\right>\right],
\eq
where $\phi_{\rm G}$ is a Gaussian distributed random field, $\left<\cdots\right>$ denotes ensemble average and the parameter $\fnl$ quantifies the departures from non-Gaussianity. The current tightest bounds on $\fnl$ come from the analysis of three-point statistics of the cosmic microwave background (CMB) by the Planck satellite, which constrain  $\fnl = -0.9 \pm 5.1\ (1\sigma)$ \cite{2019arXiv190505697P}.

The next major improvements on the precision of $\fnl$, $\sigma_{\fnl}$, are expected to come from analyses of the statistics of the galaxy distribution. For Gaussian distributed primordial fluctuations, on sufficiently large scales, the galaxy number density contrast $\delta_g$ can be written as $\delta_g(\vx) = b_1 \delta_m(\vx) + \eps(\vx)$, where $\delta_m$ denotes matter density contrast fluctuations, $\eps$ is a noise field and $b_1$ is the linear local-in-matter-density galaxy bias parameter that describes the response of galaxy number counts to long-wavelength total matter fluctuations (see Ref.~\cite{biasreview} for a review on galaxy bias). In the presence of local PNG, the galaxy distribution gets another contribution $\delta_g(\vx) \supset b_\phi \fnl \phi(\vq)$ \cite{mcdonald:2008, giannantonio/porciani:2010, PBSpaper, angulo/etal:2015, 2015JCAP...11..024A, assassi/baumann/schmidt}, where $b_\phi$ is the galaxy bias parameter that describes the response of galaxy number counts to long-wavelength primordial gravitational potential perturbations with local PNG ($\vq$ is the Lagrangian coordinate associated with the Eulerian coordinate $\vx$). This $\fnl$ contribution can therefore be used to place bounds on local PNG using galaxy observations. For example, Ref.~\cite{dalal/etal:2008} showed that the galaxy power spectrum $P_{gg}(k)$ (the Fourier transform of the two-point correlation function; $k$ denotes wavenumber) acquires a specific scale-dependent signature $\propto b_1b_\phi \fnl / k^2$ that becomes important on the largest observable scales. The galaxy bispectrum $B_{ggg}(k_1,k_2,k_3)$ (the Fourier transform of the three-point correlation function) is also a notoriously good probe of $\fnl$ \cite{scoccimarro/etal:2004, 2007PhRvD..76h3004S, jeong/komatsu:2009b, 2009PhRvD..80l3002S, 2010JCAP...07..002N, giannantonio/porciani:2010, 2011JCAP...04..006B, sefusatti/etal:2012, tellarini/etal, 2016JCAP...06..014T, 2018MNRAS.478.1341K}, and it is sensitive to it also via the non-zero primordial matter bispectrum that it induces (we discuss the $\fnl$ contributions to the power spectrum and bispectrum in more detail in Sec.~\ref{sec:model}).  Current constraints on local PNG from galaxy surveys are of order $\sigma_{\fnl} \sim 50$ \cite{slosar/etal:2008, 2011JCAP...08..033X, 2013MNRAS.428.1116R, 2014PhRvD..89b3511G, 2014PhRvL.113v1301L, 2014MNRAS.441L..16G, 2015JCAP...05..040H, 2019JCAP...09..010C}, but a number of forecast studies \cite{2008ApJ...684L...1C, 2012MNRAS.422.2854G, 2014arXiv1412.4872D, 2014arXiv1412.4671A, 2015PhRvD..91d3506F, 2015JCAP...01..042R, 2015PhRvD..92f3525A, 2015MNRAS.448.1035C, 2017PhRvD..95l3513D, 2017PDU....15...35R, 2018MNRAS.478.1341K, 2018PhRvL.121j1301C, 2019ApJ...872..126M, 2019arXiv191103964K, 2020MNRAS.492.1513G} have been suggesting that $\sigma_{\fnl} \sim \mathcal{O}(1)$ or below could be achieved with upcoming surveys.

A main theoretical uncertainty in searches for local PNG using the galaxy distribution concerns the galaxy bias parameters. They are functions of redshift and of galaxy properties such as their total mass, stellar mass, luminosity, etc. They formally describe the response of galaxy formation to the long-wavelength environment, and as a result, they are extremely challenging to predict given the many astrophysical processes involved in galaxy formation and evolution. In observational searches for local PNG, one should therefore fit for the bias parameters simultaneously with $\fnl$, but naturally, the wider the priors adopted for the bias parameters, the weaker the constraining power on $\fnl$. This motivates work on theoretical predictions of galaxy bias in order to determine the range of values a given observed galaxy sample is expected to take and/or determine relations between the various bias parameters to reduce the dimensionality of the parameter spaces explored in constraint analyses. A widely popular example of such relations is the so-called universality relation between $b_\phi$ and $b_1$. Concretely, considering gravity-only dynamics in the formation of dark matter haloes and further assuming that their mass function is universal, it can be shown that $b_\phi = 2\delta_c\left(b_1 - 1\right)$ \cite{slosar/etal:2008, 2008ApJ...677L..77M, afshordi/tolley:2008, 2010A&A...514A..46V, matsubara:2012, ferraro/etal:2012, PBSpaper, scoccimarro/etal:2012, 2017MNRAS.468.3277B}, where $\delta_c = 1.686$ is the threshold overdensity for spherical collapse. The adoption of such a relation is crucial to constrain $\fnl$. For example, in the contribution to the galaxy power spectrum $\propto b_1b_\phi \fnl / k^2$, the value of $b_1$ can be constrained by the smaller-scale part of the power spectrum, but then, if $b_\phi$ is allowed complete freedom, it becomes impossible to constrain $\fnl$ (cf.~upper left panel of Fig.~\ref{fig:contributions} below).

Effectively all recent constraint and forecast studies on $\fnl$ using the galaxy distribution adopt some form for the relation $b_\phi(b_1)$. The most popular case is the use of the universality relation mentioned above, but this relation has been shown to represent only an approximation of the bias values measured for haloes in $N$-body simulations \cite{grossi/etal:2009, desjacques/seljak/iliev:2009, 2010MNRAS.402..191P, 2010JCAP...07..013R, 2011PhRvD..84h3509H, scoccimarro/etal:2012, 2012JCAP...03..002W, baldauf/etal:2015, 2017MNRAS.468.3277B, 2020arXiv200609368B}. More specifically, the simulation results are better described by a variant of the relation $b_\phi = q 2\delta_c\left(b_1 - 1\right)$, where $q \in \left[0.5, 0.9\right]$ (the exact value depends on redshift, halo mass and halo finding criterion). Another variant of the universality relation, $b_\phi = 2\delta_c\left(b_1 - p\right)$ with $p=1.6$ was put forward by Ref.~\cite{slosar/etal:2008}, who argued that it was a more appropriate description for recently formed haloes, which could host quasars (see also Ref.~\cite{2010JCAP...07..013R}). More recently, Ref.~\cite{2020arXiv200609368B} found using hydrodynamical simulations of galaxy formation with the IllustrisTNG model \cite{Pillepich:2017jle, 2017MNRAS.465.3291W, Nelson:2018uso} that $b_\phi = 2\delta_c\left(b_1 - p\right)$ with $p \approx 0.55$ describes well the bias relation for galaxies selected by their stellar mass. Naturally, uncertainties on galaxy bias in general, and on the $b_\phi(b_1)$ relation in particular, will propagate to the resulting inferred values of $\fnl$ and $\sigma_{\fnl}$. This is a fact that has been acknowledged is past literature, but that has not been the subject of detailed and dedicated work, despite its utmost importance for PNG constraints. Our main goal in this paper is to take a few steps forward in understanding how such uncertainties can impact the inferred $\fnl$ and how to mitigate them. 

Specifically in this paper, we work with an idealized forecast setup to illustrate how galaxy bias uncertainties affect the constraints on $\fnl$ obtained with a combination of multitracer galaxy power spectrum and bispectrum data. We explore two treatments of the galaxy bias $b_\phi$: (i) one called parametrization 1, in which we assume $b_\phi = 2\delta_c\left(b_1 - p\right)$ and treat $p$ as a free parameter, and (ii) another called parametrization 2, in which instead of constraining $\fnl$, we constrain directly the product of $b_\phi\fnl$. In the latter approach, one is less interested in the actual value of $\fnl$, but more on distinguishing it from zero as $b_\phi\fnl \neq 0$ implies $\fnl \neq 0$, which is sufficient to distinguish single-field from multifield inflation. With parametrization 1, we will see that the shape of the likelihood becomes appreciably bimodal if $p$ is allowed complete freedom. This complicates the interpretation of marginalized constraints on $\fnl$, but we will see that priors $\Delta p \lesssim 0.5$ for fiducial values of $p = 0.55$ are sufficient to return unbiased constraints. Regarding parametrization 2, which is completely independent of any assumptions on the PNG bias parameters, we will find that the significance of the detection of local PNG (i.e. $\fnl \neq 0$) is similar to that from parametrization 1, when using the galaxy power spectrum. However, with parametrization 2, the addition of the galaxy bispectrum contributes much less significantly to improving the constraints, compared to the case with parametrization 1. 

Our goal in the remainder of this paper is not to draw precise quantitative estimates of $\sigma_{\fnl}$, but rather to outline and discuss some strategies to deal with galaxy bias uncertainties. These can and should be straightforwardly implemented in more complete forecast pipelines to derive more robust and survey-specific bounds on $\fnl$. The outline of this paper is as follows.  In Sec.~\ref{sec:model}, we display the expressions of the multitracer galaxy power spectrum, galaxy bispectrum and corresponding covariance matrix that we use in our analysis. In Sec.~\ref{sec:results}, we describe our forecast methodology, and present and discuss our numerical constraint results. Finally, we summarize and conclude in Sec.~\ref{sec:summary}. In App.~\ref{app:bgggderivation}, we comment on a few aspects of the derivation of the galaxy bispectrum, and in App.~\ref{app:cov}, we analyse the impact that varying levels of completeness of the calculation of the covariance matrix have on $\fnl$ constraints. 

\section{The galaxy power spectrum and bispectrum with local PNG}
\label{sec:model}

In this section we display the expressions of the galaxy power spectrum, galaxy bispectrum and corresponding covariances that we consider in our analysis.

We work with the following expression for the rest-frame galaxy number density constrast $\delta_g(\vx, z)$,
\bq\label{eq:biasexp}
\delta_g(\vx, z) &=& b_1(z)\delta_m(\vx, z) + \frac{1}{2} b_2(z) [\delta_m(\vx, z)]^2 + b_{K^2}(z)[K_{ij}(\vx, z)]^2 + \eps(\vx) + \eps_\delta(\vx)\delta_m(\vx, z) \nonumber \\
&+& \fnl \big[b_\phi(z) \phi(\vq) + b_{\phi\delta}(z)\phi(\vq)\delta_m(\vx, z) + \eps_\phi(\vx)\phi(\vq)\big],
\eq
where $\delta_m$ is a long-wavelength total matter density fluctuation, $K_{ij} = \left[\partial^i\partial^j/\nabla^2 - \delta^{ij}/3\right]\delta_m$ is a long-wavelength tidal field and $\phi$ is the long-wavelength primordial gravitational potential (these terms and their products should be understood as {\it renormalized operators} in this galaxy bias expansion \cite{mcdonald:2006, assassi/etal, PBSpaper, assassi/baumann/schmidt}). The density and tidal fields are evaluated at the evolved Eulerian position $\vx$, whereas the primordial potential $\phi$ is evaluated at the initial Lagrangian position $\vq$. The parameters $b_1, b_2, b_{K^2}, b_\phi, b_{\phi\delta}$ are the {\it galaxy bias parameters} that describe the response of the galaxy number density to the presence of the corresponding long-wavelength perturbations that each multiplies (see Ref.~\cite{biasreview} for a review); the bias parameters are a function of redshift, as well as of galaxy properties like stellar mass or luminosity. The stochastic terms $\eps, \eps_\delta, \eps_\phi$ encapsulate the dependence of galaxy number counts on the smaller-wavelength properties of the environment where the galaxies form. 

The terms in Eq.~(\ref{eq:biasexp}) are all that are needed to self-consistently derive the leading-order galaxy power spectrum and bispectrum \cite{assassi/baumann/schmidt} (see also Sec.~7 of Ref.~\cite{biasreview} for a review of the contribution of PNG to the galaxy bias expansion). For simplicity, we skip modelling redshift space distortions (RSD) \cite{2016JCAP...06..014T, 2018MNRAS.478.1341K} and always work in real space. Further, we skip considering so-called relativistic effects \cite{2011JCAP...10..031B, yoo/etal:2009, 2010PhRvD..82h3508Y, bonvin/durrer:2011, challinor/lewis:2011, 2015CQGra..32d4001J, 2020arXiv200701802W} that can affect galaxy statistics with the same scale dependence as $\fnl$. The form of their contribution is however known and it depends on the magnification and time-evolution bias parameters, which can both be estimated from the data. We note that these and other simplifications (such as ignoring observational systematics) can have an impact on the resulting $\fnl$ and $\sigma_{\fnl}$ values, but here we are more interested in the relative impact of galaxy bias uncertainties on the constraints, which depends less sensitively on the level of completion of the rest of the analysis. We retain also only terms that are linear in $\fnl$, which are the most relevant given current observational bounds, $|\fnl| \lesssim 5$ \cite{2019arXiv190505697P}.

\subsection{Multitracer galaxy power spectrum}
\label{sec:pgg}

The galaxy power spectrum $P_{gg}(k)$ is defined as $(2\pi)^3P_{gg}(k)\delta_D(\vk + \vk') = \big<\delta_g({\vk})\delta_g({\vk'})\big>$, where $\delta_g(\vk)$ is the Fourier transform of $\delta_g(\vx)$ and $\delta_D$ is the Dirac delta function (from hereon we skip writing explicitly the redshift dependence in the arguments to lighten the expressions). We make use of the multitracer framework for the galaxy power spectrum \cite{2009JCAP...10..007M, 2009PhRvL.102b1302S}, in which a given galaxy sample in a given redshift bin is split into two subsamples with different bias parameters. The multitracer technique helps to reduce sample variance errors on the largest observed scales as both galaxy samples share the same large-scale modes; this allows to efficiently constrain $\fnl$ via the relative amplitude of the clustering of the two samples. We will consider the case of two subsamples for simplicity, sample A and sample B, and consider the following estimators for their auto- and cross-power spectra
\bq\label{eq:pgg_estimators}
\hat{P}_{gg}^{\rm AA}(k) &=& \frac{1}{V_s V_k} \int_k {\rm d}^3\vk'\ \delta_g^{\rm A}(\vk')\delta_g^{\rm A}(-\vk'), \\
\hat{P}_{gg}^{\rm AB}(k) &=& \frac{1}{V_s V_k} \int_k {\rm d}^3\vk'\ \delta_g^{\rm A}(\vk')\delta_g^{\rm B}(-\vk'), \\
\hat{P}_{gg}^{\rm BB}(k) &=& \frac{1}{V_s V_k} \int_k {\rm d}^3\vk'\ \delta_g^{\rm B}(\vk')\delta_g^{\rm B}(-\vk'),
\eq
where $V_s$ is the survey volume and $\int_k {\rm d}^3\vk'$ represents integrating over a spherical shell in Fourier space with radius $k$, width $\Delta k$ and volume $V_k = 4\pi k^2\Delta_k$; the superscripts $^{\rm A,B}$ indicate the galaxy subsample, i.e., $\hat{P}_{gg}^{\rm AB}$ represents the estimator of the cross-power spectrum. Plugging the Fourier transform of Eq.~(\ref{eq:biasexp}) in the expectation value of the above equations, and retaining only terms to leading order in perturbation theory and $\fnl$ yields
\bq
\label{eq:pgg_modelAA}\big<\hat{P}_{gg}^{\rm AA}(k)\big> \equiv P^{\rm AA}_{gg}(k) &=& \left[b_1^{\rm A}\right]^2 P_{mm}(k) + 2b_1^{\rm A}b_\phi^{\rm A} \fnl P_{m\phi}(k) + P^{\rm AA}_{\eps\eps}, \\
\label{eq:pgg_modelAB}\big<\hat{P}_{gg}^{\rm AB}(k)\big> \equiv P^{\rm AB}_{gg}(k) &=& b_1^{\rm A}b_1^{\rm B} P_{mm}(k) + \left[b_1^{\rm A}b_\phi^{\rm B} + b_1^{\rm B}b_\phi^{\rm A}\right] \fnl P_{m\phi}(k), \\
\label{eq:pgg_modelBB}\big<\hat{P}_{gg}^{\rm BB}(k)\big> \equiv P^{\rm BB}_{gg}(k) &=& \left[b_1^{\rm B}\right]^2 P_{mm}(k) + 2b_1^{\rm B}b_\phi^{\rm B} \fnl P_{m\phi}(k) + P^{\rm BB}_{\eps\eps},
\eq
where $P_{mm}(k)$ denotes the linear matter power spectrum and $P_{m\phi}$ the cross matter-potential power spectrum. The linear matter density perturbations are related to the primordial potential as $\delta_m^{(1)}(\vk) = \mathcal{M}(k) \phi(\vk)$, where the superscript $^{(1)}$ indicates the linear perturbation theory contribution to the total density field $\delta_m = \delta_m^{(1)} + \delta_m^{(2)} + \cdots$, and $\mathcal{M}(k) = (2/3) k^2 T_m(k) / (\Omega_{m0} H_0^2)$, with $T_m$ being the matter transfer function, $\Omega_{m0}$ the present-day fractional cosmic mean matter density and $H_0$ the present-day Hubble expansion rate. Hence, $P_{m\phi} = P_{mm}(k)/\mathcal{M}(k)$. Finally, $P_{\eps\eps}$ is the power spectrum of the noise, which we assume to be Poissonian $P_{\eps\eps} = 1/\bar{n}_g$, where $\bar{n}_g$ is the mean observed galaxy number density; note that we ignore the cross-power spectrum of the noise fields $P^{\rm AB}_{\eps\eps}$ \cite{hamaus/etal:2010}.

\begin{figure}[t!]
	\centering
	\includegraphics[width=\textwidth]{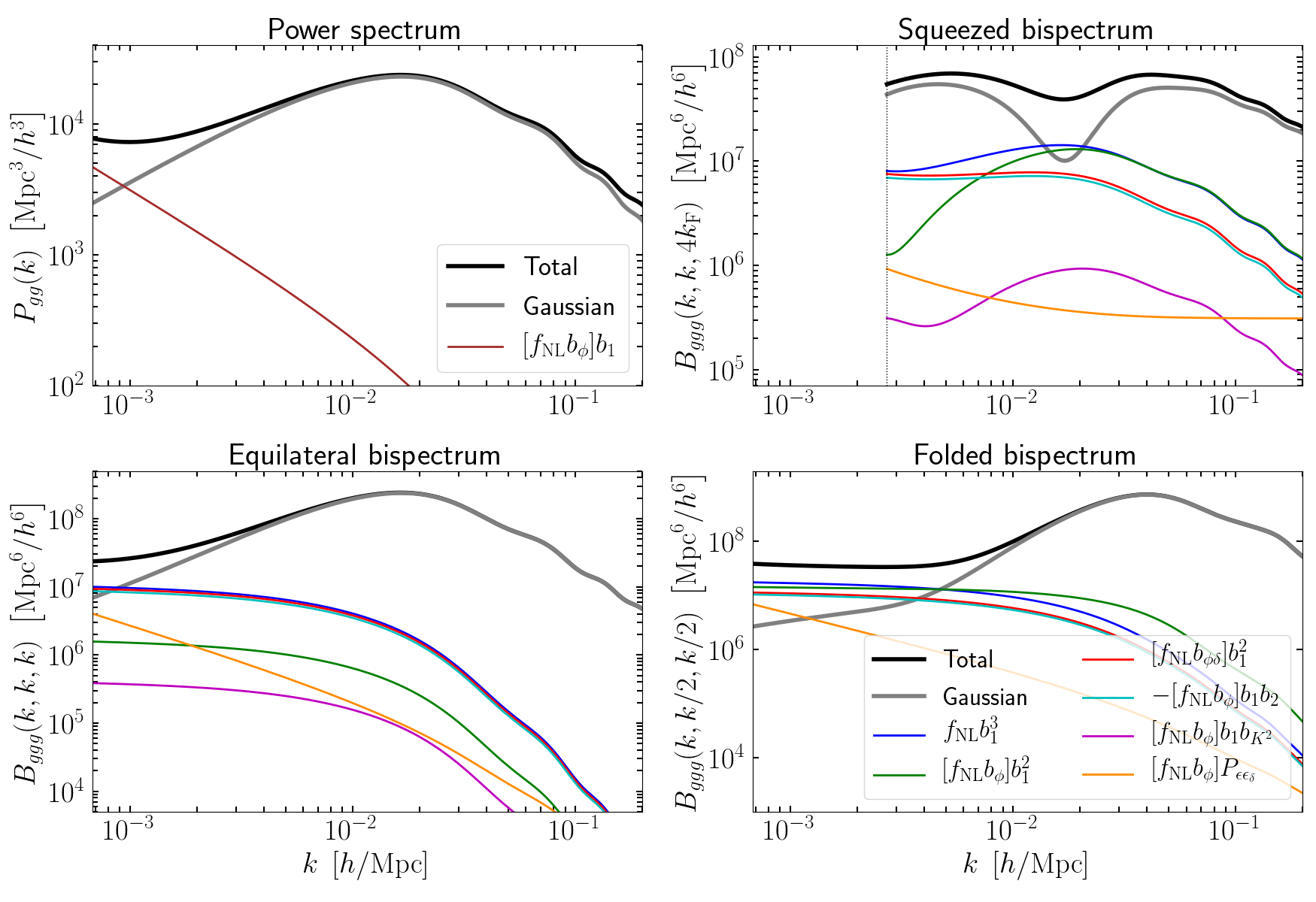}
	\caption{The different panels show, as labeled, the galaxy auto-power spectrum and the galaxy bispectrum in a squeezed configuration ($k_1=k_2 > k_3, k_3 = 4k_{\rm F} \approx 0.0027h/{\rm Mpc}$; $k_{\rm F}$ is the fundamental mode of a survey with $100 {\rm Gpc^3}/h^3$), equilateral configurations ($k_1=k_2=k_3$) and folded configurations ($k_1=2k_2=2k_3$). The result shown is the leading-order one in perturbation theory and $\fnl$ at $z=1$, and for the parameters of sample A in Tab.~\ref{table:samples}. In each panel, the black line shows the total contribution and the grey line shows the result expected for primordial Gaussian fluctuations ($\fnl = 0$). The remaining curves show the various contributions $\propto \fnl$, as labeled (the legend in the lower right panel applies to all bispectrum panels).}
\label{fig:contributions}
\end{figure}

The galaxy auto power spectrum is shown in the upper left panel of Fig.~\ref{fig:contributions}. On scales $k \lesssim 0.01\ h/{\rm Mpc}$, $T_m$ tends to a constant and the $\fnl$ contribution becomes $\propto 1/k^2$, as shown by the brown line. The ratio of the total galaxy power spectrum (black line) and the contributions assuming Gaussian primordial fluctuations (grey) is therefore scale-dependent  on large scales, which is why this signature from $\fnl$ is popularly referred to as the {\it scale-dependent bias} effect \cite{dalal/etal:2008}. 

The upper left panel of Fig.~\ref{fig:contributions} illustrates also neatly the critical importance of galaxy bias uncertainties. The local PNG signature is proportional to $b_1b_\phi\fnl$, and hence, in order to constrain $\fnl$, one needs either additional data and/or prior information on the bias parameters. If prior information on the amplitude of $P_{mm}$ is available, then the value of $b_1$ can be constrained from the smaller-scale part of the power spectrum where the $\fnl$ contribution is negligible. Nonetheless, even if $b_1$ is perfectly determined, one is still left with a perfect degeneracy between $b_\phi$ and $\fnl$. The way of breaking this degeneracy that is most popular in the literature\footnote{Note that the incorporation of more galaxy samples into the analysis does not break this degeneracy since more galaxy bias parameters would need to be added as well.} consists in relating $b_\phi$ and $b_1$ via the universality relation mentioned in the previous section, $b_\phi = 2\delta_c\left(b_1 - 1\right)$. There is however no guarantee that observed galaxies obey this relation. Indeed, as mentioned already in Sec.~\ref{sec:introduction}, studies have shown that the universality relation is not exactly satisfied even for haloes in gravity-only simulations \cite{grossi/etal:2009, desjacques/seljak/iliev:2009, 2010MNRAS.402..191P, 2010JCAP...07..013R, 2011PhRvD..84h3509H, scoccimarro/etal:2012, 2012JCAP...03..002W, baldauf/etal:2015, 2017MNRAS.468.3277B, 2020arXiv200609368B}. Further, $b_\phi = 2\delta_c\left(b_1 - 1.6\right)$ has been argued to be a better description of haloes that have recently undergone a major merger \cite{slosar/etal:2008, 2010JCAP...07..013R}, and $b_\phi = 2\delta_c\left(b_1 - 0.55\right)$ has been put forward as a more adequate description of simulated stellar-mass selected galaxies \cite{2020arXiv200609368B}. These results suggest that the exact $b_\phi(b_1)$ relation is very likely tracer-dependent (i.e., different for different galaxy types) and currently fairly uncertain. This makes it a pressing matter to study and develop ways to incorporate and mitigate uncertainties around the $b_\phi(b_1)$ relation in $\fnl$ constraints; that is the subject of Sec.~\ref{sec:results} below.


\subsection{Galaxy bispectrum}
\label{sec:bggg}

The galaxy bispectrum is defined as $(2\pi)^3 B_{ggg}(k_1, k_2, k_3)\delta_D({\vk_{123}}) = \big<\delta_g({\vk_1})\delta_g({\vk_2})\delta_g({\vk_3})\big>$ (in our notation, $\vk_{123} = \vk_1 + \vk_2 + \vk_3$), and we consider the following estimator,
\bq\label{eq:bggg_estimator}
\hat{B}_{ggg}(k_1,k_2,k_3) = \frac{1}{V_sV_{123}} \int_{k_1} {\rm d}^3\vk_a \int_{k_2} {\rm d}^3\vk_b \int_{k_3} {\rm d}^3\vk_c \ \delta_g(\vk_a)\delta_g(\vk_b)\delta_g(\vk_c)\delta_D(\vk_{abc}),
\eq
where $V_{123} = 8\pi^2k_1k_2k_3\Delta k_1\Delta k_2\Delta k_3$. In our analysis we work with the bispectrum of only one of the samples used in the multitracer power spectrum part for simplicity. We use the bispectrum of one of the samples rather than the two samples combined to reduce the number of free bias parameters. In App.~\ref{app:bgggderivation} we comment on a few aspects of the derivation of the galaxy bispectrum (see also Refs.~\cite{jeong/komatsu:2009b,2009PhRvD..80l3002S,  2010JCAP...07..002N, giannantonio/porciani:2010, 2011JCAP...04..006B, sefusatti/etal:2012, tellarini/etal, assassi/baumann/schmidt, 2016JCAP...06..014T, 2018MNRAS.478.1341K} for a number of past works on the galaxy bispectrum in local PNG cosmologies). The final result can be written as
\bq\label{eq:bggg_model}
\big<\hat{B}_{ggg}(k_1,k_2,k_3)\big> \equiv B_{ggg}(k_1,k_2,k_3) = B_{ggg}^{\rm G}(k_1,k_2,k_3) + B_{ggg}^{\rm NG}(k_1,k_2,k_3),
\eq
where 
\bq\label{eq:bggg_model_G}
&&B_{ggg}^{\rm G}(k_1,k_2,k_3) = b_1^3B_{mmm}(k_1, k_2, k_3) + \big[2 b_1 P_{mm}(k_1)P_{\eps\eps_\delta} + {\rm (2\ perm.)}\big] + B_{\eps\eps\eps} \nonumber \\
&+&\Big[b_1^2b_2P_{mm}(k_1)P_{mm}(k_2) + {\rm (2\ perm.)}\Big] + \Big[2b_1^2b_{K^2}\left(\mu_{12}^2 - \frac13\right)P_{mm}(k_1)P_{mm}(k_2) + {\rm (2\ perm.)}\Big] \nonumber \\
\eq
is the contribution due to nonlinear gravitational clustering that is present even if $\fnl = 0$, and 
\bq\label{eq:bggg_model_NG}
&&B_{ggg}^{\rm NG}(k_1,k_2,k_3) = \Bigg[ 2b_1^3\fnl\frac{P_{mm}(k_1)P_{mm}(k_2)}{\mathcal{M}(k_1)\mathcal{M}(k_2)}\mathcal{M}(k_3) + 2b_\phi\fnl\frac{P_{mm}(k_1)}{\mathcal{M}(k_1)}P_{\eps\eps_\delta} \nonumber \\
&+&  b_1^2b_\phi\fnl P_{mm}(k_1)P_{mm}(k_2) \Bigg(\mu_{12}\bigg(\frac{k_1}{k_2\mathcal{M}(k_1)} + \frac{k_2}{k_1\mathcal{M}(k_2)}\bigg) + 2F_2(k_1, k_2, \mu_{12})\bigg(\frac{1}{\mathcal{M}(k_1)} + \frac{1}{\mathcal{M}(k_2)}\bigg)\Bigg) \nonumber \\
&+&  \bigg(b_1^2b_{\phi\delta}  + b_1b_2b_{\phi} + 2b_1b_{K^2}b_{\phi} \Big(\mu_{12}^2 - \frac13\Big)\bigg)\fnl P_{mm}(k_1)P_{mm}(k_2) \bigg(\frac{1}{\mathcal{M}(k_1)} + \frac{1}{\mathcal{M}(k_2)}\bigg) + {\rm (2\ perm.) \Bigg]} \nonumber \\
\eq
is the contribution due to local PNG that is $\propto \fnl$.\footnote{In the forecast analysis on PNG presented in Ref.~\cite{2018MNRAS.478.1341K}, the authors include also the contribution from a 1-loop term $\propto \fnl$  given by an integral of a certain trispectrum configuration (Fourier transform of the four-point function). This term was first derived and discussed in Refs.~\cite{jeong/komatsu:2009b, 2009PhRvD..80l3002S}, although in the context of a galaxy bias expansion that did not include the $\phi$ field in Eq.~(\ref{eq:biasexp}). In this paper, we work at tree level in the galaxy bispectrum, and importantly, with a set of operators in the galaxy bias expansion that is complete and closed under renormalization \cite{assassi/baumann/schmidt}. Contributions from 1-loop terms on the large scales of interest are therefore absorbed by renormalized bias parameters at tree level.} In the equations above, $\mu_{ab}$ is the cosine of the angle between the $k_a$ and $k_b$ legs of the triangle, and likewise to the case of the power spectrum, we work to leading order in the bispectrum (i.e., second order in perturbation theory) and in $\fnl$. Further, $B_{\eps\eps\eps}$ and $P_{\eps\eps_\delta}$ denote, respectively, the bispectrum of the noise field $\eps$ and the cross-power spectrum of the fields $\eps$ and $\eps_\delta$. Assuming Poissonian noise these are given by $B_{\eps\eps\eps} = 1/\bar{n}_g^2$ and $P_{\eps\eps_\delta} = b_1/(2\bar{n}_g)$. To derive the second term on the right-hand side of Eq.~(\ref{eq:bggg_model_NG}), we have also assumed Poissonian statistics to relate $P_{\eps\eps_\phi} = (b_\phi/b_1)P_{\eps\eps_\delta}$. Further, $B_{mmm}$ denotes the tree-level matter bispectrum given by
\bq\label{eq:bmmm}
B_{mmm}(k_1, k_2, k_3) = 2F_2(\vk_1, \vk_2) P_{mm}(k_1)P_{mm}(k_2) + {\rm (2\ perm.)},
\eq
where
\bq\label{eq:f2}
F_2(\vk_1, \vk_2) \equiv F_2(k_1, k_2, \mu_{12}) = \frac57 + \frac{\mu_{12}}{2}\left[\frac{k_1}{k_2} + \frac{k_2}{k_1}\right] + \frac27 \mu_{12}^2
\eq
is the second-order perturbation theory mode-coupling kernel \cite{Bernardeau/etal:2002}.

In Fig.~\ref{fig:contributions}, the upper right, lower left and lower right panels show, respectively, the galaxy bispectrum for squeezed ($k_1=k_2 > k_3\approx0.0027h/{\rm Mpc}$), equilateral ($k_1=k_2=k_3$) and folded ($k_1=2k_2=2k_3$) configurations. A first noteworthy and well-known aspect is that the local PNG contributions are most prominent in the squeezed-limit. This is shown by the larger difference between the black and grey lines in the upper right panel. Note that the local PNG contribution is sizeable in the squeezed bispectrum for all modes $k_1=k_2$ probed, which is why increasing the maximum wavenumber analysed $k_{\rm max}$ can help improve significantly $\fnl$ constraints with the bispectrum, but less so with the power spectrum. Further, contrary to the power spectrum case, in the galaxy bispectrum the $\fnl$ contribution is not perfectly degenerate with $b_\phi$. This degeneracy is broken by the terms $\propto b_1^3 \fnl$ (blue lines in Fig.~\ref{fig:contributions}) and $\propto b_1^2b_{\phi\delta}\fnl$ (red lines in Fig.~\ref{fig:contributions}). In the latter case, $b_{\phi\delta}$ can also be calculated assuming the universality of the halo mass function:
\bq\label{eq:univbphidelta}
b_{\phi\delta} = b_\phi - b_1 + 1 + \delta_c[b_2 - (8/21)(b_1-1)]. 
\eq
The same caveats about the validity of the universality relation for $b_\phi$ apply likewise to $b_{\phi\delta}$. To the best of our knowledge, however, this $b_{\phi\delta}$ expression has never been tested neither for halos in gravity-only simulations nor simulated galaxies in hydrodynamical simulations. In our forecasts below, when we study deviations from the universality relation for $b_\phi$, we shall still assume the universality relation for $b_{\phi\delta}$. The latter should be eventually tested with $N$-body simulations, but we note that this term contributes a smaller amount compared to terms $\propto b_\phi$ to the constraining power on $\fnl$, and hence, uncertainties on $b_{\phi\delta}$ are not as critical. 

\subsection{Covariance of the multitracer galaxy power spectrum and bispectrum}
\label{sec:covs}

In our forecast analysis we work with the following Gaussian likelihood function
\bq\label{eq:likelihood}
\mathcal{L}(\vtheta)  \propto {\rm exp} \Big[-\frac12\left(\vM(\vtheta) - \vD\right)^t{\Cov}^{-1}\left(\vM(\vtheta) - \vD\right)\Big], 
\eq
where $\vD$ is the data vector, ${\Cov}$ its covariance matrix and $\vM$ is a theoretical prediction for the data vector that depends on parameters $\vtheta$. The data vector consists of hypothetical measurements of the real-space multitracer power spectrum and bispectrum
\bq\label{eq:datavector}
\vD = \Big\{ \hat{P}_{gg}^{\rm AA} , \hat{P}_{gg}^{\rm AB} , \hat{P}_{gg}^{\rm BB} , \hat{B}_{ggg}^{\rm AAA}\Big\};
\eq
we will work with the bispectrum of subsample A, which is the higher number density one (cf. Sec.~\ref{sec:methodology}). Our theoretical predictions $\vM$ are evaluated using Eqs.~(\ref{eq:pgg_modelAA}), (\ref{eq:pgg_modelAB}), (\ref{eq:pgg_modelBB}), (\ref{eq:bggg_model}), (\ref{eq:bggg_model_G}) and (\ref{eq:bggg_model_NG}); we construct the data vector using the same equations evaluated at a fiducial cosmology. The covariance matrix of the data vector contains three main blocks:

\begin{equation}\label{eq:cov_def}
{\Cov} = 
\begin{pmatrix}
\Cov^{PP}  & \Cov^{BP} \\[1.5ex]
\cdots  & \Cov^{BB} \\[1.5ex]
\end{pmatrix}
\,\,,
\end{equation}
where $\cov^{PP}$, $\cov^{BP}$ and $\cov^{BB}$ denote, respectively, the covariance of the power spectrum part of the data vector, the cross-covariance of the bispectrum and power spectrum, and the covariance of the bispectrum part; the dots indicate the block is equal to the corresponding symmetric block. In App.~\ref{app:cov}, we outline the derivation of the covariance matrix, and comment on the relative impact of the various contributions to it (see also Refs.~\cite{2006PhRvD..74b3522S, 2012A&A...540A...9M, 2013MNRAS.429..344K, 2019MNRAS.482.4883C, 2009A&A...508.1193J, 2018PhRvD..97d3532C, 2017PhRvD..96b3528C, 2019MNRAS.490.4688R, sqbcov} for works on bispectrum covariances). In this section, we limit ourselves to displaying only the final expressions. For $\cov^{PP}$, we consider only the contribution from the disconnected part of the four-point function (the so-called Gaussian term); for the case of the multitracer power spectrum, the result is a $3\times3$ block diagonal matrix:
\begin{equation}\label{eq:covPP}
{\Cov}^{PP}(k_1, k_2) = 2 \frac{(2\pi)^3\delta_{k_1k_2}}{V_sV_{k_1}}
\begin{pmatrix}
[P_{gg}^{AA}(k_1)]^2 & P_{gg}^{AB}(k_1)P_{gg}^{AA}(k_1) & [P_{gg}^{AB}(k_1)]^2 \\[1.5ex]
\cdots & \frac12\left[P_{gg}^{AA}(k_1)P_{gg}^{BB}(k_1) + [P_{gg}^{AB}(k_1)]^2\right] & P_{gg}^{BB}(k_1)P_{gg}^{AB}(k_1) \\[1.5ex]
\cdots & \cdots & [P_{gg}^{BB}(k_1)]^2
\end{pmatrix}
\,\,.
\end{equation}
We evaluate the bispectrum covariance block as $\cov^{BB} = \cov^{BB}_{PPP} + \cov^{BB}_{BB}$, with (in our notation, the superscripts in $\cov$ indicate which estimators we are taking the covariance of, whereas the subscripts label the various contributions\footnote{The labels $B$ here should not be confused with the labels ${\rm B}$ that refer to one of the multitracer subsamples.})
\bq\label{eq:covBBPPP}
\cov^{BB}_{PPP}(k_1, k_2, k_3,k_1', k_2', k_3') = \delta_{TT'} \frac{(2\pi)^6S_{\rm shape}}{V_s V_{123}} P_{gg}(k_1)P_{gg}(k_2)P_{gg}(k_3)
\eq
and
\bq\label{eq:covBBBB}
\cov^{BB}_{BB}(k_1, k_2, k_3,k_1', k_2', k_3') &=& \frac{(2\pi)^3}{V_s}\frac{U(k_1, k_1')}{V_{123}V_{1'2'3'}} B_{ggg}(k_1', k_2, k_3)B_{ggg}(k_1, k_2', k_3') \delta_{k_1k_1'} + {\rm (8\ perm.)}, \nonumber \\
\eq
where $U(k_1, k_1') = 16\pi^3 k_2k_3k_2'k_3' \Delta k_2 \Delta k_3 \Delta k_2' \Delta k_3' \Delta k_1$, $\delta_{TT'}$ is a Kronecker delta function that is non-zero only if both triangles $T = \{k_1, k_2, k_3\}$ and $T' = \{k_1', k_2', k_3'\}$ are the same, and $S_{\rm shape} = 6,2,1$ for equilateral, isosceles and scalene triangles, respectively; in Eq.~(\ref{eq:covBBBB}), the permutations are all 9 that link each side of one triangle to each side of the other. Finally, in our main results we will neglect the cross-covariance term, $\cov^{BP} = 0$; its contribution is analysed and discussed in App.~\ref{app:cov}.

\section{Results}
\label{sec:results}

In this section we present our main results on the impact of local PNG galaxy bias uncertainties on $\fnl$ constraints. We begin in Sec.~\ref{sec:methodology} by describing our idealized forecast setup and the two galaxy bias parametrizations that we explore in this paper. The numerical forecast results are then shown and discussed in Secs.~\ref{sec:numerics1} and \ref{sec:numerics2}.

\subsection{Methodology and forecast setup}
\label{sec:methodology}
\begin{table*}
\centering
\begin{tabular}{lcccccccccccccc}
\toprule
& {\rm Sample} & $\bar{n}_g$\ $\left[{h^3/{\rm Mpc^3}}\right]$ & $b_1$ & $b_2$ & $b_{K^2}$ &  $b_{\phi}$ & $b_{\phi\delta}$ & \\
\midrule
& A & $1.74\times10^{-3}$ & $1.58$ & $-0.62$ & $-0.17$ &  $3.49$ & $1.48$ & \\
& B & $1.07\times10^{-4}$ & $2.37$ & $0.66$ & $-0.39$ &  $6.15$ & $5.01$ & \\
\bottomrule
\end{tabular}
\caption{Mean number density and bias parameters of the two galaxy subsamples A and B used in the multitracer power spectrum part of the analysis. In the bispectrum part, we consider the bispectrum of the higher number density subsample A. The samples are further assumed to have a mean redshift of $z = 1$ and span a volume in the Universe of $V_s = 100{\rm Gpc^3}/h^3$.}
\label{table:samples}
\end{table*}

In our forecast study we consider a galaxy sample at redshift $z = 1$. For the multitracer power spectrum part, we split this galaxy sample into two subsamples A and B as follows. We assume the galaxy sample follows a stellar mass function described by a Schechter function \cite{1976ApJ...203..297S}
\bq\label{eq:smf}
\Phi(M_*) = \phi_* \left(\frac{M_*}{M_*'}\right)^{-\alpha} {\rm exp}\big[-M_* / M'\big],
\eq
with $\phi_* = 0.0074\ h^3/{\rm Mpc^3}/{\rm dex}$, $M_*' = 10^{11}\ M_{\odot}/h$ and $\alpha = 0.38$; we have verified this is sufficiently realistic in that it matches well the stellar mass function of the IllustrisTNG simulations at $z=1$ \cite{Pillepich:2017jle, 2017MNRAS.465.3291W, Nelson:2018uso}. We then define the two subsamples via a cut in stellar mass $M_*$: subsample A with $M_* \in \left[5\times10^{10} ; 2\times10^{11}\right]M_{\odot}/h$ and subsample B with $M_* > 2\times10^{11}M_{\odot}/h$. The mean number density $\bar{n}_g$ and linear bias parameter $b_1$ of each sample are obtained by integrating, respectively, 
\bq\label{eq:ng}
\bar{n}_g = \int {\rm dln}M_* \frac{\Phi(M_*)}{{\rm ln}10}\ \ ,\ \ b_1 = \frac{1}{\bar{n}_g}\int {\rm dln}M_* \frac{\Phi(M_*)}{{\rm ln}10} b_1^{\rm T}(M_*)
\eq
over the corresponding mass ranges. In the equation for $b_1$, $b_1^{\rm T}(M_*)$ denotes the linear halo bias fitting formula of Ref.~\cite{2010ApJ...724..878T} evaluated in terms of stellar mass according to the prescription described in Ref.~\cite{2020arXiv200609368B}. We evaluate the higher-order bias terms $b_2, b_{K^2}$ using, respectively, the polynomial fit of Ref.~\cite{lazeyras/etal} $b_2 = 0.412 - 2.143b_1 + 0.929b_1^2 + 0.008b_1^3$, and the Lagrangian linear-in-matter-density prediction $b_{K^2} = -(2/7)(b_1-1)$. The fiducial values of $b_\phi$ are obtained with the variant of the universality relation $b_\phi = 2\delta_c(b_1-p)$ with $p = 0.55$, which is roughly the value preferred by the stellar-mass selected objects in the IllustrisTNG model \cite{2020arXiv200609368B}. We use Eq.~(\ref{eq:univbphidelta}) to evaluate $b_{\phi\delta}$. These specifications of the two subsamples are summarized in Tab.~\ref{table:samples}. For the bispectrum part of the analysis, we consider the bispectrum of subsample A, which has higher number density.\footnote{We have also ran constraints using the bispectrum of the full combined sample and found the same conclusions.}

For the fiducial cosmology we assume a spatially flat $\Lambda{\rm CDM}$ model with present-day baryon density $\Omega_{b0} = 0.0486$, cold dark matter density $\Omega_{c0} = 0.2603$, dimensionless Hubble parameter $h = 0.6774$, scalar spectral index $n_s = 0.967$ and primordial scalar amplitude parameter $\mathcal{A}_s = 2.068\times10^{-9}$ at a pivot scale $k_{\rm pivot} = 0.05\ /{\rm Mpc}$. We neglect the impact of neutrino masses, $\Omega_{m0} = \Omega_{b0} + \Omega_{c0}$. We further consider a fiducial value of $\fnl=3$, which is within the range currently allowed by CMB observations $\fnl = -0.9 \pm 5.1\ (1\sigma)$ \cite{2019arXiv190505697P}; we choose a non-zero value for $\fnl$ to entertain the possibility of non-zero detections in our analysis, which is the most interesting scenario. For simplicity and to remain focused on the impact of galaxy bias uncertainties, we will vary only the cosmological parameters $\fnl$ and $\mathcal{A}_s$. In real data analysis, one would always adopt priors on the remaining cosmological parameters from other data sets anyway. We evaluate all the linear power spectra and transfer functions with the {\tt CAMB} code \cite{camb, 2011ascl.soft02026L}.

Our assumed survey volume is $V_s = 100{\rm Gpc^3}/h^3$, which is roughly the comoving volume of a full-sky survey spanning $z  = 1 \pm 0.5$. We consider 30 $k$ bins between $k_{\rm min} = k_{\rm F} = \pi/V_s^{1/3}$ and $k_{\rm max} = 0.2\ h/{\rm Mpc}$ uniformly distributed in log-space. This value of $k_{\rm max}$ is approximately that above which our leading-order treatment of the galaxy power spectrum and bispectrum becomes inadequate. In power spectrum only analysis (denoted by $P_{gg}$-only from hereon), the value of $k_{\rm max}$ is not as important as in combined analysis with the bispectrum (denoted by $P_{gg}+B_{ggg}$ from hereon); this is because the latter can still be appreciably sensitive to $\fnl$ on small scales for squeezed configurations with large $k_1, k_2$ and small $k_3$ (cf.~upper right panel of Fig.~\ref{fig:contributions}). We have checked that with the more conservative option $k_{\rm max} = 0.1\ h/{\rm Mpc}$, the $P_{gg}+B_{ggg}$ analysis had effectively the same constraining power on $\fnl$ as the $P_{gg}$-only analysis. We therefore opted for a higher $k_{\rm max}$ value to study the consequences of galaxy bias uncertainties when the constraining power of the data is stronger. With this $k$-binning, we evaluate the bispectrum at all possible triangles with $k_{\rm min} \leq k_1, k_2, k_3 \leq k_{\rm max}$, which yields a total of $873$ triangles. 

In our numerical results, we discuss the impact of galaxy bias uncertainties on local PNG constraints under the following two characterizations of the parameter space:

\begin{itemize}
\item \underline{Parametrization 1: $b_\phi = 2\delta_c\left(b_1 - p\right)$.} In this case, $b_\phi$ is related to $b_1$ via the variant of the universality relation $b_\phi = 2\delta_c\left(b_1 - p\right)$, with $p$ being treated as a free parameter. We will be specifically interested on the impact of different assumed priors on $p$. The parameter space that we wish to constrain is 11 dimensional and given by
\bq\label{eq:params1}
\vtheta = \{\fnl, b_1^{\rm A}, b_1^{\rm B}, P_\eps^{\rm A}, P_\eps^{\rm B}, p, \delta_{\mathcal{A}_s}, b_2^{\rm A}, b_{K^2}^{\rm A}, P_{\eps\eps_\delta}^{\rm A} , B_{\eps\eps\eps}^{\rm A} \},
\eq
where $\delta_{\mathcal{A}_s}$ is defined as $\mathcal{A}_s = \mathcal{A}_{s, \rm fidu}\left[1 + \delta_{\mathcal{A}_s}\right]$, with $\mathcal{A}_{s, \rm fidu}$ being the fiducial value. For $P_{gg}$-only analyses we vary just the first six of these parameters; we keep $\delta_{\mathcal{A}_s}$ fixed, which would otherwise be perfectly degenerate with $b_1^{\rm A}, b_1^{\rm B}$. Note that we assume the value of $p$ to be the same for both subsamples in the multitracer analysis. Important to recall is also the fact that we still assume the universality prediction of Eq.~(\ref{eq:univbphidelta}) to derive $b_{\phi\delta}$, although this term contributes subdominantly compared to others.

\item \underline{Parametrization 2: $\fnl b_\phi$.} In this case, instead of attempting to make assumptions on $b_\phi$ to constrain $\fnl$, one fits instead directly for the products of $\fnl b_\phi$ and $\fnl b_{\phi\delta}$ that contribute to the galaxy power spectrum and bispectrum. The parameter space then becomes 13 dimensional and given by
\bq\label{eq:params2}
\vtheta = \{[\fnl b_\phi^{\rm A}], [\fnl b_\phi^{\rm B}], b_1^{\rm A}, b_1^{\rm B}, P_\eps^{\rm A}, P_\eps^{\rm B}, \fnl, \delta_{\mathcal{A}_s}, [\fnl b_{\phi\delta}^{\rm A}], b_2^{\rm A}, b_{K^2}^{\rm A}, P_{\eps\eps_\delta}^{\rm A} , B_{\eps\eps\eps}^{\rm A} \}.
\eq
For $P_{gg}$-only analyses we vary just the first six of these parameters; this parametrization is completely independent of any assumptions on $b_\phi$, but one loses the ability to constraint the absolute value of $\fnl$. The value of this approach lies instead on the possibility to {\it model-independently} detect non-zero $[\fnl b_\phi^{\rm A}]$ or $[\fnl b_\phi^{\rm B}]$, which would imply $\fnl \neq 0$, and consequently, rule out single-field models of inflation. For $P_{gg}+B_{ggg}$ analysis, the absolute value of $\fnl$ can still be formally constrained because of the term $\propto b_1^3\fnl$ in Eq.~(\ref{eq:bggg_model_NG}) (although with much reduced constraining power, as we will see in Sec.~\ref{sec:numerics2}).\footnote{Explicit bounds on $\fnl$ can also always be obtained with parametrization 2 by imposing priors on $b_\phi$ {\it a posteriori}.}
\end{itemize}
Finally, we sample the parameter space with the {\tt EMCEE} {\tt Python} implementation \cite{2013PASP..125..306F} of the affine-invariant Markov Chain Monte Carlo (MCMC) sampler proposed in Ref.~\cite{2010CAMCS...5...65G}. We assume wide uninformative linear priors on all parameters and use 32 {\it walkers} with a nominal chain convergence criterion that the chain size must be $100$ times the autocorrelation time and the latter having varied less than $1\%$ the last time it was calculated (every few thousand samples, in our case). We have also visually inspected the contour plots and found them to be stable enough for our conclusions well before nominal convergence was achieved. We have further rerun the constraints with the walkers initialized at different starting points, which returned the same constraints. As another cross-check, we have further compared the outcome of {\tt EMCEE} with that of a two-dimensional grid search for the parameters $\fnl$ and $p$ in parametrization 1 (which as we will see next displays a non-trivial bimodal shape of the likelihood), which we found to be effectively indistinguishable. 

\subsection{Results from parametrization 1: $b_\phi = 2\delta_c\left(b_1 - p\right)$}
\label{sec:numerics1}

\begin{figure}[t!]
	\centering
	\includegraphics[width=\textwidth]{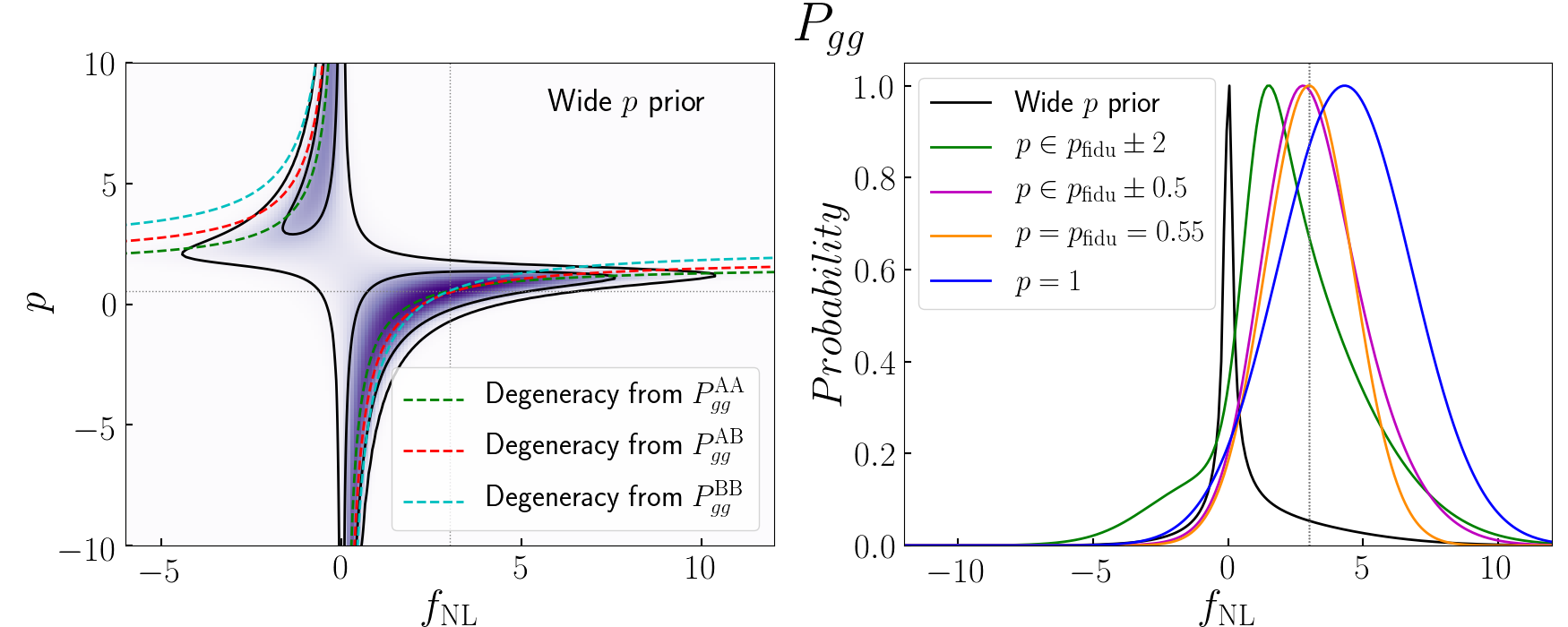}
	\caption{Constraints on $\fnl$ and $p$ from $P_{gg}$-only analysis and keeping all the remaining parameters fixed at the fiducial values. The left panel shows the shape of the likelihood function (shaded purple) and the $1\sigma$ and $2\sigma$ confidence contour levels (black lines) for the case without assumed priors on $p$. The dashed lines mark the $\fnl-p$ degeneracy directions of the spectra in the multitracer analysis, as labeled (cf.~Eqs.~(\ref{eq:degdirs})). The right panel shows the marginalized constraints on $\fnl$ for different assumed priors on $p$, as labeled. The dotted lines indicate the fiducial parameter values. }
\label{fig:intuition_0_Pgg}
\end{figure}

To build intuition for the shape of the constraints on $\fnl$ with the $b_\phi = 2\delta_c\left(b_1 - p\right)$ parametrization, we study first  the results from constraints in which only $\fnl$ and $p$ are varied and the remaining parameters are kept fixed at their fiducial values. In Fig.~\ref{fig:intuition_0_Pgg}, this is shown for the $P_{gg}$-only constraints. The left panel shows that the two-dimensional distribution is appreciably bimodal with one peak at $\fnl > 0$ and $p < b_1^A, b_2^A$, and another at $\fnl < 0$ and $p > b_1^A, b_2^A$. This is as expected since the multitracer galaxy power spectrum is weakly sensitive to a simultaneous change of sign of $\fnl$ and $b_1-p$; for a single galaxy sample, the power spectrum would be completely unchanged under this change of sign. This shape of the distribution can be understood by inspecting the degeneracy directions from the three spectra in our multitracer power spectrum analysis. Concretely, Eqs.~(\ref{eq:pgg_modelAA}), (\ref{eq:pgg_modelAB}) and (\ref{eq:pgg_modelBB}) remain unchanged if $\fnl$ and $p$ vary, respectively, according to the relations
\bq\label{eq:degdirs}
{\rm Degeneracy\ from\ } P_{gg}^{\rm AA}\ \ ::\ \ p &=& b_{1, \rm fidu}^{\rm A} - \frac{{\fnl}_{,\rm fidu} (b_{1, \rm fidu}^{\rm A} - p_{\rm fidu})}{\fnl}, \nonumber \\
{\rm Degeneracy\ from\ } P_{gg}^{\rm AB}\ \ ::\ \ p &= & \frac{2b_{1, \rm fidu}^{\rm A}b_{1, \rm fidu}^{\rm B}}{b_{1, \rm fidu}^{\rm A}+b_{1, \rm fidu}^{\rm B}} - \frac{{\fnl}_{,\rm fidu} (2 - \frac{b_{1, \rm fidu}^{\rm A}+b_{1, \rm fidu}^{\rm B}}{b_{1, \rm fidu}^{\rm A}b_{1, \rm fidu}^{\rm B}}p_{\rm fidu})}{\fnl}, \nonumber \\
{\rm Degeneracy\ from\ } P_{gg}^{\rm BB}\ \ ::\ \ p &=& b_{1, \rm fidu}^{\rm B} - \frac{{\fnl}_{,\rm fidu} (b_{1, \rm fidu}^{\rm B} - p_{\rm fidu})}{\fnl},
\eq
where the subscript $_{\rm  fidu}$ indicates the fiducial value. These relations are shown by the dashed lines in the left panel of Fig.~\ref{fig:intuition_0_Pgg}. All three degeneracy directions asymptote to $p \to \pm \infty$ as $\fnl \to 0$. Hence, given that the likelihood is not negligibly small at $\fnl = 0$, this implies that $p$ cannot be constrained in our $P_{gg}$ analysis.\footnote{In our ``wide $p$ prior'' results, we still enforce $p \in \left[-100, +100\right]$ to prevent the chains from indefinitely exploring the degeneracy.} On the other hand, an increase in $\fnl$ can be compensated by $p$ values that approach $b_1^{\rm A}$ and $b_1^{\rm B}$. Crucially, however, due to the different values of $b_1$ of the two subsamples, this compensation is not perfect and $\fnl$ can be constrained: in our concrete idealized setup, the two-dimensional $1\sigma$ contours span $-2 \lesssim \fnl \lesssim 7$. 

It follows also from the above discussion that single-tracer power spectrum analyses cannot constrain $\fnl$ if $p$ is varied freely. Similarly, multitracer analyses with the subsamples allowed to take on different values of $p$ cannot also constrain $\fnl$, unless prior information is added on the $p$ parameters. This has important ramifications to galaxy selection strategies for multitracer analyses. Here, we have assumed $p$ to be constant for samples selected by stellar-mass, which is in accordance with the results obtained in Ref.~\cite{2020arXiv200609368B} using simulations with the IllustrisTNG model. However, the combination of samples with varying selection criteria is also a perfectly viable option (say, a stellar-mass selected sample A and active galactic nuclei as sample B), but which forcibly requires assuming different $p$ values for the various samples. We do not investigate this scenario explicitly in this paper (which is nonetheless qualitatively similar to cases discussed next with priors on $p$), although we stress that uncertainties around the $b_\phi(b_1)$ relation should be a factor to take into consideration in the design of multitracer analyses.

\begin{figure}[t!]
	\centering
	\includegraphics[width=\textwidth]{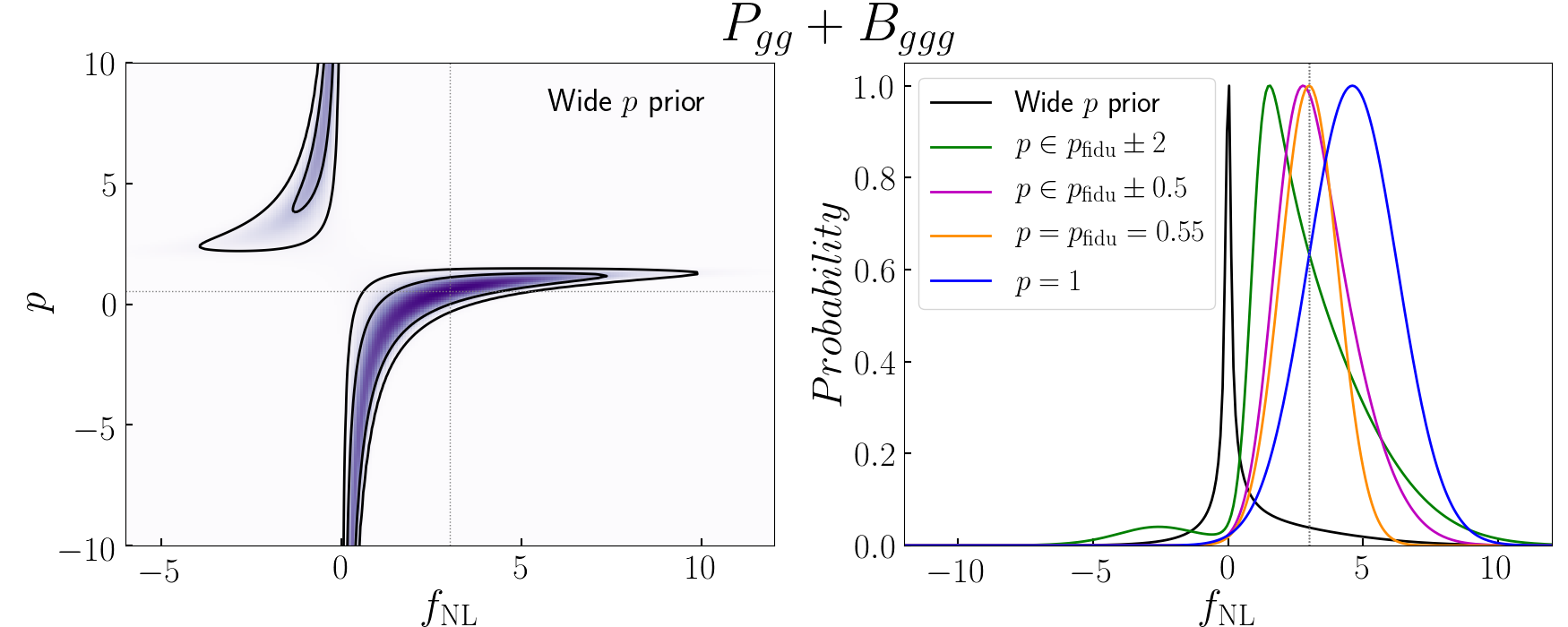}
	\caption{Same as Fig.~\ref{fig:intuition_0_Pgg}, but for combined $P_{gg} + B_{ggg}$ analyses.}
\label{fig:intuition_0_PggBggg}
\end{figure}

The bimodal shape of the two-dimensional constraints on $\fnl$ and $p$ demands special care in the interpretation of the corresponding one-dimensional marginalized constraints. The latter are shown for $\fnl$ by the black solid line in the right panel of Fig.~\ref{fig:intuition_0_Pgg}. The curve is strongly peaked around $\fnl = 0$ since the degeneracy directions that leave $p$ unconstrained assign most of the volume of the distribution there (the volume is not simply infinite because our wide prior $p \in \left[-100, +100\right]$ makes it finite). The $1\sigma$ range around the maximum of this distribution excludes even the fiducial value $\fnl = 3$. The green and magenta lines in the right panel of Fig.~\ref{fig:intuition_0_Pgg} show the result from constraints assuming stronger priors on $p$: $p \in p_{\rm fidu} \pm 2$ and $p \in p_{\rm fidu} \pm 0.5$, respectively. As expected, the stronger priors reduce the volume of the distribution around $\fnl = 0$, which progressively centers the marginalized constraints around the fiducial $\fnl$ value. We have also explicitly checked (not shown) that a prior $p \in p_{\rm fidu} \pm 0.15$ yields effectively the same constraints as those from keeping $p$ fixed at the fiducial value (orange line in the right panel of Fig.~\ref{fig:intuition_0_Pgg}).

\begin{figure}[t!]
	\centering
	\includegraphics[width=\textwidth]{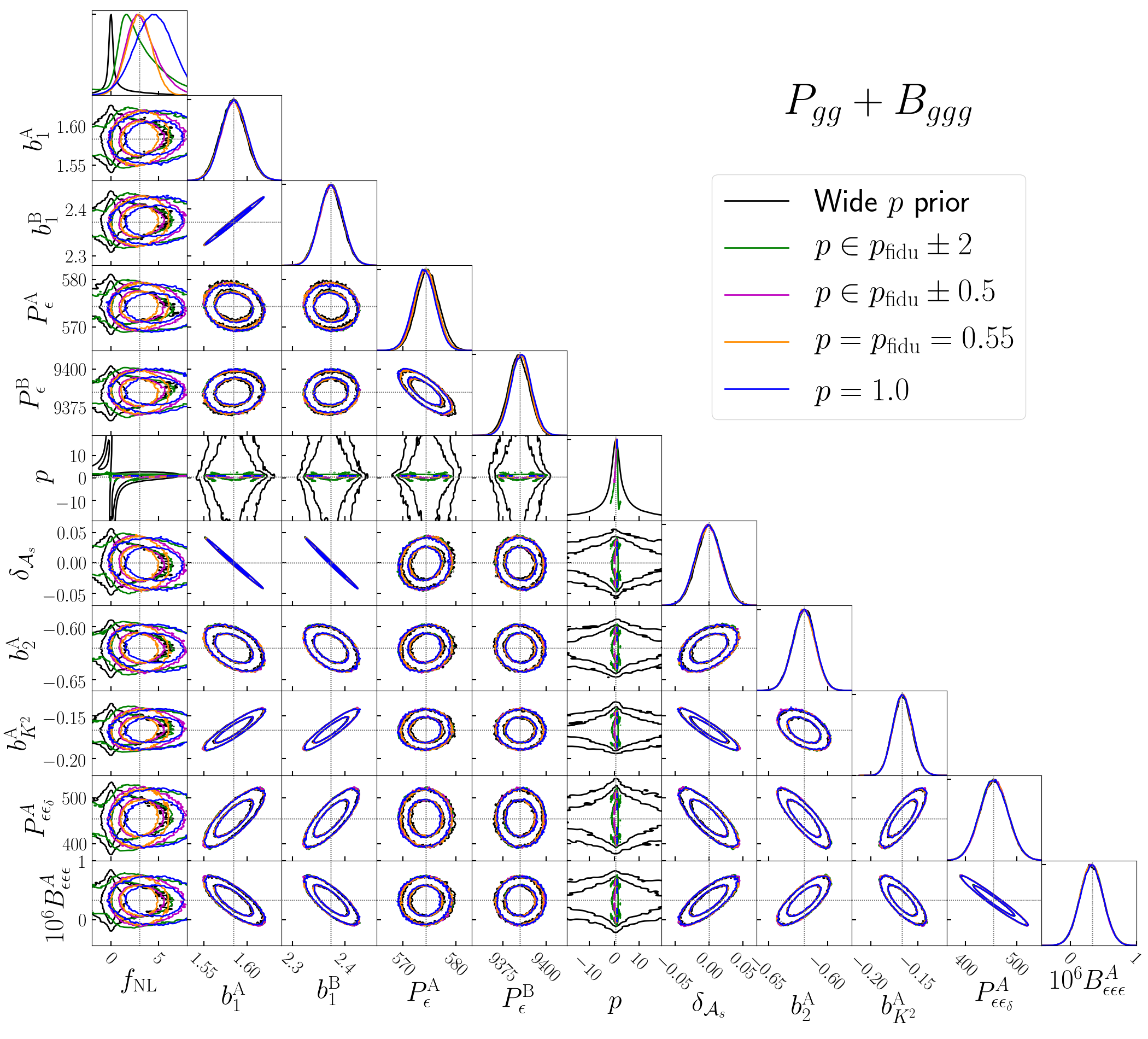}
	\caption{Triangle constraints plot from combined $P_{gg}+B_{ggg}$ analyses with the parametrization 1, $b_\phi = 2\delta_c(b_1-p)$. The panels in the diagonal show one-dimensional marginalized constraints and those in the off-diagonal show the $1\sigma$ and $2\sigma$ confidence levels of two-dimensional marginalized constraints. The different colors indicate different assumed priors on $p$, as labeled (only distinguishable along the $\fnl$ direction; the different priors on $p$ also naturally have an impact along the $p$ direction, but which are only hardly visible on the scale of the plot). The values of $P_\eps^{\rm A}$, $P_\eps^{\rm B}$, $P_{\eps\eps_\delta}^{\rm A}$ have dimensions of power spectra and $B_{\eps\eps\eps}^{\rm A}$ of bispectra.}
\label{fig:triangle_0_PggBggg}
\end{figure}

Figure \ref{fig:intuition_0_PggBggg} shows the same as Fig.~\ref{fig:intuition_0_Pgg}, but for the combined $P_{gg}+B_{ggg}$ analysis. The same qualitative features are present and the only differences are the quantitatively tighter constraints provided by the bispectrum information. In particular, both the $1\sigma$ and $2\sigma$ contours are disjoint in the $P_{gg}+B_{ggg}$ case, but in the $P_{gg}$-only case, the $1\sigma$ contours are the only disjoint ones.

It is also interesting to discuss the result depicted by the blue line in the right panels of Figs.~\ref{fig:intuition_0_Pgg} and \ref{fig:intuition_0_PggBggg}, which shows the constraints from keeping $p$ fixed but equal to the wrong value $p=1$ (that of the universality relation). The two main consequences of choosing $p=1$ when the fiducial is $p=0.55$ are (i) a broadening of the distribution because the values of $b_\phi$ become smaller and this reduces the constraining power; and (ii) a shift in the position of the maximum of the distribution towards higher $\fnl$. In our idealized forecast setup, the corresponding $1\sigma$ bounds still encompass our fiducial value of $\fnl = 3$, but the result illustrates nonetheless the potential dangers of keeping $p$ fixed to a wrong value (see e.g.~Refs.~\cite{slosar/etal:2008, 2015JCAP...05..040H, 2019JCAP...09..010C} for examples of constraint studies where two different values of $p$ are used, $p=1$ and $p=1.6$).

The discussion above for the case in which only $\fnl$ and $p$ are varied in the constraints facilitates the interpretation of the results when all parameters in Eq.~(\ref{eq:params1}) are free. These are shown by the triangle constraints plot of Fig.~\ref{fig:triangle_0_PggBggg}. A first point to note is that the constraints on all parameters except $\fnl$ and $p$ are insensitive to the different priors on $p$. Further, the figure displays also the well known degeneracy between $\mathcal{A}_s$ and the galaxy bias parameter $b_1$, which is perfect in power spectrum analysis, but it is broken by the bispectrum information. The constraints on the $\fnl-p$ plane are similar to those discussed already in Figs.~\ref{fig:intuition_0_Pgg} and \ref{fig:intuition_0_PggBggg}, although just slightly looser because of small degeneracies that arise with the remaining parameters. 

The left panel of Fig.~\ref{fig:sigmas} summarizes the marginalized $1\sigma$ bounds on $\fnl$ for $P_{gg}$-only and combined $P_{gg} + B_{ggg}$ analyses, and for different priors on $p$ with the parametrization 1, $b_\phi = 2\delta_c(b_1-p)$. As discussed above, when $p$ is allowed to vary freely, both $P_{gg}$-only and $P_{gg} + B_{ggg}$ cases yield biased $1\sigma$ constraints that do not encompass the fiducial value (the $2\sigma$ intervals do encompass $\fnl = 3$; not shown). The prior $p \in p_{\rm fidu} \pm 0.5$ is however sufficient to return $1\sigma$ constraints that are nearly indistinguishable from those obtained assuming perfect knowledge of the $p$ parameter of the galaxies. Also as discussed above, the adoption of the universality relation value $p=1$ shifts the distributions slightly upwards and degrades slightly the constraining power of the data: the net effect are constraints on $\fnl$ shifted from the fiducial value by $\approx 0.3\sigma$  and $\approx 0.7\sigma$ for the $P_{gg}$-only and combined $P_{gg}+B_{ggg}$ analyses, respectively. 

\subsection{Results from parametrization 2: $\fnl b_\phi$}
\label{sec:numerics2}

\begin{figure}
  \subfigure
  {
    \includegraphics[width=0.5\textwidth]{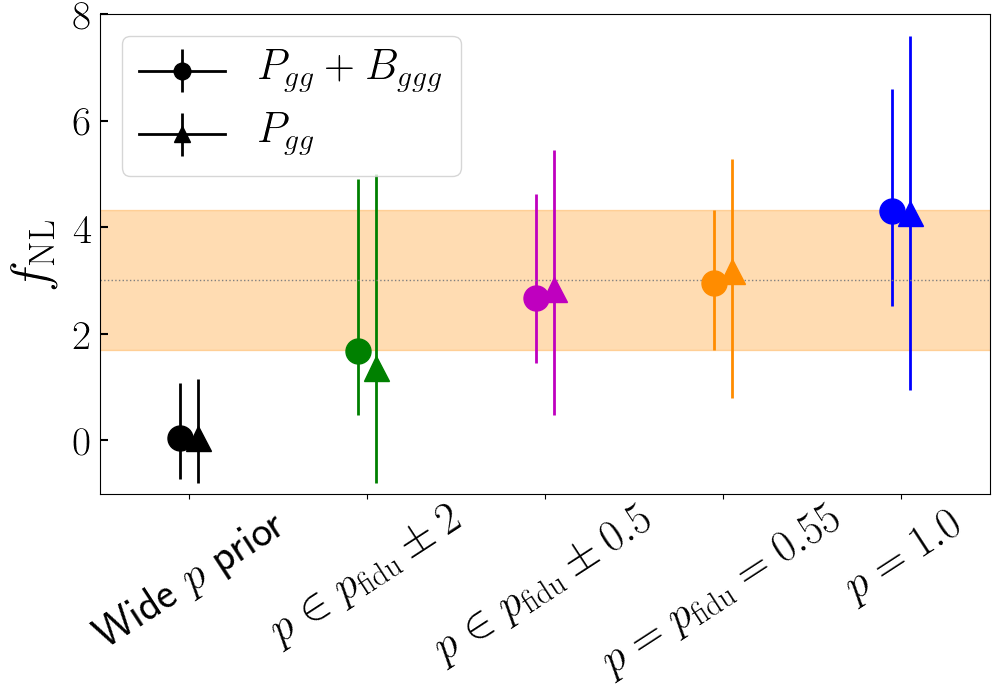}
  }
  \subfigure
  {
    \includegraphics[width=0.5\textwidth]{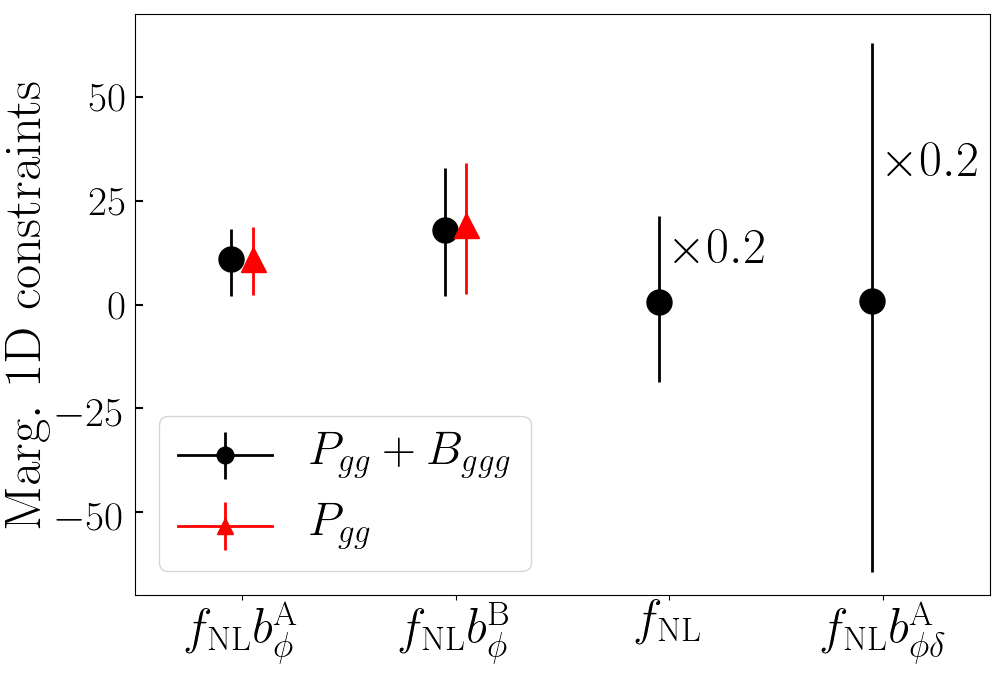}
  }
  \caption{(Left panel) Summary of $1\sigma$ marginalized constraints on $\fnl$ with parametrization 1, $b_\phi = 2\delta_c(b_1-p)$, for both $P_{gg}$-only and combined $P_{gg}+B_{ggg}$ analyses, and as a function of different assumed priors on $p$, as labeled. The orange band marks the $P_{gg} + B_{ggg}$ constraints with $p = p_{\rm fidu} = 0.55$. (Right panel) Summary of $1\sigma$ marginalized constraints on $\fnl b_\phi^{\rm A}$, $\fnl b_\phi^{\rm B}$, $\fnl$ and $\fnl b_{\phi\delta}^{\rm A}$ with parametrization 2, for both $P_{gg}$-only and combined $P_{gg}+B_{ggg}$ analyses, as labeled. The constraints on $\fnl$ and $\fnl b_{\phi\delta}^{\rm A}$ are scaled down by a factor of $5$ to improve visualization.}
  \label{fig:sigmas}
\end{figure}

\begin{figure}[t!]
	\centering
	\includegraphics[width=\textwidth]{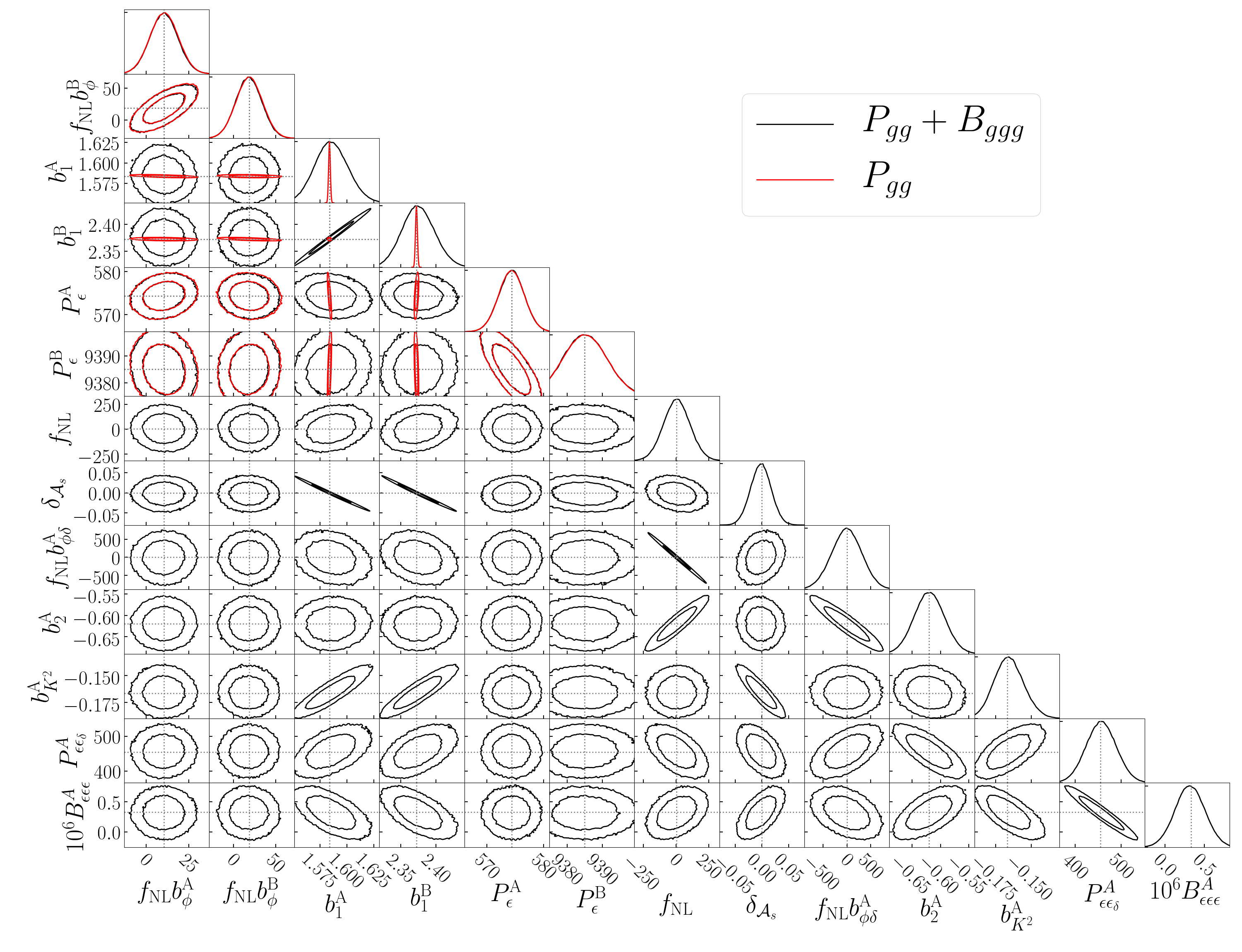}
	\caption{Same as Fig.~\ref{fig:triangle_0_PggBggg}, but for the $P_{gg}$-only and combined $P_{gg}+B_{ggg}$ analysis with parametrization 2, $\fnl b_\phi$. The width of the constraints on $b_1^{\rm A}$ and $b_1^{\rm B}$ in the $P_{gg}$ case is barely noticeable in the scale of the plot. Recall, in the $P_{gg}$-only case, we do not treat $\mathcal{A}_s$ as a free parameter.}
\label{fig:triangle_1_PggBggg}
\end{figure}

We turn our attention now to the constraints obtained under parametrization 2 (cf.~Eq.~(\ref{eq:params2})). These are shown in the triangle constraints plot of Fig.~\ref{fig:triangle_1_PggBggg}; the right panel of Fig.~\ref{fig:sigmas} summarizes the one-dimensional marginalized $1\sigma$ bounds on the four parameters relevant to the detection of local PNG. 

In the $P_{gg}$-only case, our idealized setup would be able to distinguish $\fnl b_\phi^{\rm A}$ and $\fnl b_\phi^{\rm B}$ from zero with approximately $1.25\sigma$ and $1.14\sigma$ significance, respectively. This is in line with the significance of approximately $1.30\sigma$ with which the $P_{gg}$-only analysis would be able to distinguish non-zero $\fnl$ in parametrization 1 with perfect knowledge of the $b_\phi(b_1)$ relation (cf.~orange triangle symbol in the left panel of Fig.~\ref{fig:sigmas}). This is as expected and indicates that, in multitracer power spectrum analyses, detections of non-zero $\fnl$ that completely circumvent assumptions on the $b_\phi(b_1)$ relation are possible and retain the same significance as when assuming perfect knowledge on the galaxy bias parameters; the only price to pay is that the exact value of $\fnl$ is left undetermined.

Interestingly, even though unfortunately, the addition of the bispectrum information does not tighten the constraints on $\fnl b_\phi^{\rm A}$ and $\fnl b_\phi^{\rm B}$ (cf.~comparable red and black symbols in the right panel of Fig.~\ref{fig:sigmas}). This is in contrast with the tightening of the constraints on $\fnl$ that the addition of the bispectrum attains with parametrization 1 (cf.~circle vs.~triangle symbols in the left panel of Fig.~\ref{fig:sigmas}). The reason behind the lack of constraining power of the bispectrum on local PNG with parametrization 2 can be traced back to a strong degeneracy that arises between $\fnl$ and $\fnl b_{\phi\delta}$, as shown in the corresponding panel of Fig.~\ref{fig:triangle_1_PggBggg}. Concretely, in parametrization 1, the constraining power of the bispectrum on local PNG comes from the six terms $\propto \fnl$ shown in Fig.~\ref{fig:contributions}. On the other hand, in the case of parametrization 2, the same number of terms exists, but the terms $\propto b_1^3\fnl$ and $\propto b_1^2 [\fnl b_{\phi\delta}]$ are not constrained by the power spectrum part of the data vector and are therefore free to compensate each others' effects on the bispectrum. This degeneracy is in fact quite strong, given the similar scale dependence that these terms display on small-scales in squeezed configurations (cf.~blue and red curves in the upper right panel of Fig.~\ref{fig:contributions}), which is where most of the constraining power lies. This works to effectively reduce the contribution from local PNG to the galaxy bispectrum and make it less constraining.\footnote{For such large values of $\fnl$, terms $\propto \fnl^2$ that we neglected in the galaxy bispectrum become more important and can impact the constraints we obtained with parametrization 2. The point still stands however that the galaxy bispectrum has reduced importance with parametrization 2 because of the strong degeneracy between $\fnl$ and $\fnl b_{\phi\delta}$.}

Note however that the addition of bispectrum information is always useful in general to break the degeneracy between the $b_1$ parameters and $\mathcal{A}_s$; the latter is assumed fixed at the fiducial value in our $P_{gg}$-only analysis. This does not have however a particularly strong impact on the local PNG constraints, which are only very weakly correlated with $b_1$ (cf.~the similar constraints on $\fnl$ from $P_{gg}$-only and $P_{gg}+B_{ggg}$ analyses, despite markedly different constraints on $b_1$ in Fig.~\ref{fig:triangle_1_PggBggg}).

\section{Summary \& Conclusions}
\label{sec:summary}

One of the main open questions in cosmology concerns the degree of non-Gaussianity of the distribution of the energy fluctuations in the primordial Universe. The simplest single-field models of inflation predict Gaussian distributed fluctuations, and hence, any detection of primordial non-Gaussianity (PNG) would rule out these simpler models and open the door to more elaborate multifield constructions. The next major breakthroughs on observational searches for PNG are expected to come from analyses of the statistics of the large-scale galaxy distribution and are focused on so-called local-type PNG, parametrized by the amplitude of $\fnl$ in Eq.~(\ref{eq:fnl}). The tightest current bounds from the CMB set $\fnl = -0.9 \pm 5.1\ (1\sigma)$  \cite{2019arXiv190505697P}, but upcoming large-scale structure surveys are expected to be able to probe $\sigma_{\fnl} \sim \mathcal{O}(1)$ \cite{2008ApJ...684L...1C, 2012MNRAS.422.2854G, 2014arXiv1412.4872D, 2014arXiv1412.4671A, 2015PhRvD..91d3506F, 2015JCAP...01..042R, 2015PhRvD..92f3525A, 2015MNRAS.448.1035C, 2017PhRvD..95l3513D, 2017PDU....15...35R, 2018MNRAS.478.1341K, 2018PhRvL.121j1301C, 2019ApJ...872..126M, 2019arXiv191103964K, 2020MNRAS.492.1513G}.

A major challenge in using galaxies to constrain cosmology concerns the uncertainties associated with the bias parameters of the observed galaxy samples. In the case of local PNG, these parameters give rise to strong degeneracies with $\fnl$ that drastically reduce the constraining power of the data. The most popular way of breaking these degeneracies involves establishing relations between the most relevant galaxy bias parameters, like the relation between the galaxy bias parameters $b_1$ and $b_\phi$ (cf.~Eq.~(\ref{eq:biasexp})). In the context of gravity-only dynamics and assuming universality of the halo mass function, these two bias parameters are related by $b_\phi = 2\delta_c\left(b_1 - p\right)$ with $p=1$. There is, however, no reason to expect this to be a good description for actually observed galaxies. In fact Refs.~\cite{slosar/etal:2008, 2010JCAP...07..013R} showed that the $b_\phi(b_1)$ relation is sensitive to the formation time of the host haloes (with $p=1.6$ being a more adequate description of recent mergers), and Ref.~\cite{2020arXiv200609368B} showed that $p=0.55$ provides a better description of stellar mass selected galaxies simulated with the IllustrisTNG model. This uncertainty on the $b_\phi(b_1)$ relation will invariably impact the bounds on $\fnl$. Our goal in this paper was to take a few steps in the direction of determining which strategies are available to make $\fnl$ constraints more robust to such galaxy bias uncertainties.

In this paper, we worked with an idealized forecast setup with galaxies at $z=1$ in a volume $V_s = 100{\rm Gpc}^3/h^3$ (cf.~Sec.~\ref{sec:methodology}) to analyse the impact that different assumptions on $b_\phi$ and its relation to $b_1$ can have on local PNG constraints. We have considered both multitracer power spectrum data (called $P_{gg}$-only), as well as its combination with the galaxy bispectrum (called $P_{gg} + B_{ggg}$). We have focused on two parametrizations: (i) parametrization 1, in which $b_\phi = 2\delta_c\left(b_1 - p\right)$ and $p$ is treated as a free parameter, and (ii) parametrization 2, in which one fits directly for products of $\fnl$ and the local PNG bias parameters. The latter parametrization bypasses the need for any assumptions on the $b_\phi(b_1)$ relation, but makes it harder to pin down the exact value of $\fnl$\footnote{Makes it impossible, in fact, for power spectrum only analysis.}. Nonetheless, it can still be useful to distinguish $\fnl$ from zero, and therefore, discriminate between single-field and multifield inflation. Our main findings can be summarized as follows:

\begin{itemize}
\item The constraints on the $\fnl-p$ plane are bimodal and the marginalized bounds on $\fnl$ depend sensitively on the assumed priors on $p$ (cf.~Figs.~\ref{fig:intuition_0_Pgg} and \ref{fig:intuition_0_PggBggg}). In our setup, priors $p \in p_{\rm fidu} \pm 0.5$ can yield constraints similar to those with $p$ fixed to the fiducial value $p_{\rm fidu} = 0.55$ (cf.~Fig.~\ref{fig:sigmas}). 

\item Fixing $p$ to the wrong value affects both $\fnl$ and $\sigma_{\fnl}$. In our setup for a fiducial value $p=0.55$, the adoption of the universality relation $p=1$ in the combined $P_{gg}+B_{ggg}$ analysis shifts the marginalized constraints on $\fnl$ upwards by $\approx 0.7\sigma$ (cf.~Fig.\ref{fig:sigmas}).

\item In $P_{gg}$-only analysis, the significance of the detection of $\fnl \neq 0$ with parametrization 1 and $\fnl b_\phi \neq 0$ with parametrization 2 is the same (cf.~Fig.~\ref{fig:sigmas}). This shows that the detection of local PNG using galaxies could be made completely independently of the $b_\phi(b_1)$ relation.

\item The addition of the galaxy bispectrum improves constraints with parametrization 1, but not with parametrization 2 because the sensitivity of the bispectrum to local PNG is reduced by a strong degeneracy that arises between $\fnl$ and $[\fnl b_{\phi\delta}]$ (cf.~discussion in Sec.~\ref{sec:numerics2}).

\end{itemize}

Our work shows that constraints on $\fnl$ can depend critically on the assumed $b_\phi(b_1)$ relation, which motivates further theoretical studies on it. Such an interesting set of studies can involve determining the values of $p$ predicted from current state-of-the-art galaxy formation simulations (as Ref.~\cite{2020arXiv200609368B} did recently with IllustrisTNG) and use the mean and scatter of the predictions to inform priors on $p$. This is akin to the case of baryonic effects on the small-scale total matter power spectrum that is a major source of uncertainty in weak-lensing data analysis, and whose priors are also often informed by hydrodynamical simulations \cite{2015MNRAS.454.1958M, 2019OJAp....2E...4C, 2020JCAP...04..019S, 2020JCAP...04..020S, 2020MNRAS.495.4800A}. 

It is important to note also that $b_\phi = 2\delta_c(b_1 - p)$ is a purely phenomenological variant of the universality relation that was shown to describe well the stellar-mass selected objects simulated with IllustrisTNG for $p=0.55$. It could well be the case that the same functional form is not a good fit to the results from other galaxy formation simulations, nor to galaxy samples selected by properties other than stellar-mass. In fact, the results of Ref.~\cite{2020arXiv200609368B} show that $b_\phi = 2\delta_c(b_1 - p)$ ceases to be an adequate description for objects selected by color or black hole accretion rate. These are particularly relevant considerations for multitracer analyses with samples selected by different criteria, in which case it would be likely inadequate to assume a fixed form of $b_\phi(b_1)$ for all samples (cf.~discussion in Sec.~\ref{sec:numerics1}). These remarks all motivate further work to determine better functional forms for the $b_\phi(b_1)$ relation and/or design alternative ways to place priors on $b_\phi$. Likewise, it would be interesting to test also the validity of the universality relation Eq.~(\ref{eq:univbphidelta}) for the second-order bias parameter $b_{\phi\delta}$.

\acknowledgments
We would like to thank Giovanni Cabass, Elisabeth Krause and Fabian Schmidt for useful comments and discussions. The author acknowledges support from the Starting Grant (ERC-2015-STG 678652) ``GrInflaGal'' from the European Research Council.

\appendix

\section{Aspects of the derivation of the galaxy bispectrum of Eq.~(\ref{eq:bggg_model})}
\label{app:bgggderivation}

The steps to derive Eq.~(\ref{eq:bggg_model}) involve plugging the Fourier transform of Eq.~(\ref{eq:biasexp}) into the expectation value of $\big<\delta_g(\vk_a)\delta_g(\vk_b)\delta_g(\vk_c)\big>$ (which appears in the expectation value of Eq.~(\ref{eq:bggg_estimator})), and then simply retain leading-order terms in perturbation theory, as well as terms $\propto \fnl$. This is largely a tedious straightforward exercise, although there are a few steps that are more subtle than others. We display a few of these here to illustrate some typical derivation steps and refer the interested reader to Refs.~\cite{giannantonio/porciani:2010, 2011JCAP...04..006B, 2016JCAP...06..014T, assassi/baumann/schmidt, 2018JCAP...12..035D} for other recent detailed calculations of galaxy bispectra in $\fnl$ cosmologies.

\underline{\it Example 1} 

Let us consider as a first example a term involving the tidal field $K_{ij}(\vx)$. Specifically, the term $\propto b_1 b_{K^2} b_\phi \fnl$ in Eq.~(\ref{eq:bggg_model_NG}) follows from
\bq\label{eq:example_a_1}
b_1b_{K^2}b_\phi \fnl \Bigg[\int_{\vp}\big<\delta_m(\vk_a)K_{ij}(\vp)K_{ij}(\vk_b - \vp)\frac{\delta_m^{(1)}(\vk_c)}{\mathcal{M}(k_c)}\big> + \left(\vk_b \leftrightarrow \vk_c\right)\Bigg] + {\rm (2\ perm.)},
\eq
where $\int_{\vp} \equiv \int {\rm d}^3\vp / (2\pi)^3$ and we have used already that $\phi(\vk) = \delta_m^{(1)}(\vk)/\mathcal{M}(k)$. In Fourier space, $K_{ij}(\vk) = \left(k_ik_j/k^2 - \delta_{ij}/3\right)\delta_m(\vk)$, and the derivation proceeds as
\bq\label{eq:example_a_2}
&&b_1b_{K^2}b_\phi \fnl \Bigg[\int_{\vp}\frac{1}{\mathcal{M}(k_c)}\left(\frac{p_ip_j}{p^2} - \frac{\delta_{ij}}{3}\right)\left(\frac{(k_{b,i}-p_i)(k_{b,j}-p_j)}{|\vk_b-\vp|^2} - \frac{\delta_{ij}}{3}\right)\big<\delta_m(\vk_a)\delta_m(\vp)\delta_m(\vk_b-\vp)\delta_m^{(1)}(\vk_c)\big> \nonumber \\
&& + \left(\vk_b \leftrightarrow \vk_c\right)\Bigg] + {\rm (2\ perm.)} \nonumber \\
&& = b_1b_{K^2}b_\phi\fnl \Bigg[\int_{\vp}\frac{1}{\mathcal{M}(k_c)}\left(\frac{p_ip_j}{p^2} - \frac{\delta_{ij}}{3}\right)\left(\frac{(k_{b,i}-p_i)(k_{b,j}-p_j)}{|\vk_b-\vp|^2} - \frac{\delta_{ij}}{3}\right)  \nonumber \\
&&\ \ \ \ \ \ \ \ \ \ \ \ \ \ \ \ \ \ \times \Big((2\pi)^6 P_{mm}(k_a)P_{mm}(k_c) \Big[\delta_D(\vk_{ab} - \vp)\delta_D(\vk_c + \vp) + \delta_D(\vk_{cb} - \vp)\delta_D(\vk_a + \vp)\Big] \nonumber \\
&&\ \ \ \ \ \ \ \ \ \ \ \ \ \ \ \ \ \ + \big<\delta_m(\vk_a)\delta_m(\vp)\delta_m(\vk_b-\vp)\delta_m^{(1)}(\vk_c)\big>_c\Big) \nonumber \\
&&\ \ \ \ \ \ \ \ \ \ \ \ \ \ \ \ \ \ + \left(\vk_b \leftrightarrow \vk_c\right)\Bigg] + {\rm (2\ perm.)} \nonumber \\ 
&& = 2b_1b_{K^2}b_\phi\fnl (2\pi)^3 \Bigg[\left(\mu_{ac}^2 - \frac13\right) \frac{P_{mm}(k_a)P_{mm}(k_c)}{\mathcal{M}(k_c)} + \left(\mu_{ab}^2 - \frac13\right) \frac{P_{mm}(k_a)P_{mm}(k_b)}{\mathcal{M}(k_b)}\Bigg]\delta_D(\vk_{abc}) + {\rm (2\ perm.)} \nonumber \\
&& = 2b_1b_{K^2}b_\phi\fnl (2\pi)^3 \left(\mu_{ab}^2 - \frac13\right) P_{mm}(k_a)P_{mm}(k_b) \Bigg[\frac{1}{\mathcal{M}(k_a)} + \frac{1}{\mathcal{M}(k_b)}\Bigg]\delta_D(\vk_{abc}) + {\rm (2\ perm.)},
\eq
where in the first equality we have used Wick's theorem to write the four-point function in terms of the product of two two-point functions and the connected four-point function; in the second equality we have integrated over $\vp$ using the Dirac delta functions and dropped the connected four-point function contribution, which is negligible on the scales of interest; finally, in the third equality we have reshuffled the permutations to write down the expression in a more economic manner. Plugging this expression into the expectation value of Eq.~(\ref{eq:bggg_estimator}) (and further assuming sufficiently narrow bins that it becomes a good approximation to skip the angular and bin-averages) yields the corresponding term in Eq.~(\ref{eq:bggg_model_NG}). 

\underline{\it Example 2} 

As another example, consider the term $\propto b_1^3\fnl$ in Eq.~(\ref{eq:bggg_model_NG}), which comes from the contribution of $\fnl$ to the linear matter density field
\bq\label{eq:example_b_1}
\delta_m^{(1)}(\vk) = \mathcal{M}(k) \bigg[\phi_{\rm G}(\vk) + \fnl \int_{\vp} \phi_{\rm G}(\vp) \phi_{\rm G}(\vk - \vp) \bigg].
\eq
This term then follows from $b_1^3\langle\delta_m(\vk_a)\delta_m(\vk_b)\delta_m(\vk_c)\rangle$ as 
\bq\label{eq:example_b_2}
&& b_1^3 \mathcal{M}(k_c)\bigg<\delta_m(\vk_a)\delta_m(\vk_b)\int_{\vp}\bigg(\phi_{\rm G}(\vk_c) + \fnl \int_{\vp} \phi_{\rm G}(\vp) \phi_{\rm G}(\vk_c - \vp)\bigg)\bigg> + {\rm (2\ perm.)} \nonumber \\ 
&=& b_1^3 \fnl \mathcal{M}(k_a)\mathcal{M}(k_b)\mathcal{M}(k_c)\int_{\vp} \big<\phi_{\rm G}(\vk_a)\phi_{\rm G}(\vk_b) \phi_{\rm G}(\vp) \phi_{\rm G}(\vk_c - \vp)\big> + {\rm (2\ perm.)} \nonumber \\ 
&=& b_1^3 \fnl \mathcal{M}(k_a)\mathcal{M}(k_b)\mathcal{M}(k_c)\int_{\vp} (2\pi)^6 P_{\phi\phi}(k_a)P_{\phi\phi}(k_b)\big[\delta_D(\vk_a+\vp)\delta_D(\vk_{bc} - \vp) + \delta_D(\vk_b+\vp)\delta_D(\vk_{ac} - \vp)\big] \nonumber \\  
&+& {\rm (2\ perm.)} \nonumber \\
&=& 2 b_1^3 \fnl (2\pi)^3 \frac{P_{mm}(k_a)P_{mm}(k_b)}{\mathcal{M}(k_a)\mathcal{M}(k_b)} \mathcal{M}(k_c) \delta_D(\vk_{abc})  + {\rm (2\ perm.)},
\eq
where in the first equality we have used that $\delta_m^{(1)}(\vk) = \mathcal{M}(k) \phi(\vk)$ and discarded the bispectrum of $\phi_{\rm G}$ (which is zero for a Gaussian field) and next-to-leading-order contributions; in the second equality we have used Wick's theorem to write the four-point function as the product of two two-point functions (and discarded the connected four-point function, which is zero); finally, in the third equality we have integrated over $\vp$ using the Dirac delta functions and have written the auto power spectrum of the primordial potential as $P_{\phi\phi}(k) = P_{mm}(k) / \mathcal{M}(k)^2$. In keeping with the narrow bin approximation mentioned above, plugging this expression in the expectation value of Eq.~(\ref{eq:bggg_estimator}) yields the desired term in Eq.~(\ref{eq:bggg_model_NG}).

\underline{\it Example 3} 

Another perhaps more subtle term is that $\propto b_1^2 b_\phi \fnl$ in Eq.~(\ref{eq:bggg_model_NG}). For this term, it matters explicitly to leading order that the primordial potential $\phi$ in Eq.~(\ref{eq:biasexp}) is evaluated at the Lagrangian position $\vq$, which is related to the potential evaluated at the Eulerian position $\vx$ via a displacement term as
\bq\label{eq:example_c_1}
\phi_{\rm Lag}(\vq) = \phi_{\rm Eul}(\vx) - s^i(\vx) \partial_i\phi_{\rm Eul}(\vx),
\eq
where the displacement field is given by $s^i(\vx) = -(\partial_i/\nabla^2) \delta^{(1)}(\vx)$. In Fourier space, we have\footnote{Here, we used the convolution theorem and that $i(k_i/k^2)\delta_m^{(1)}(\vk)$ and $ik_i\phi_{\rm Eul}(\vk)$ are the Fourier transforms of $s^i(\vx)$ and $\partial_i\phi_{\rm Eul}(\vx)$, respectively; the imaginary unit $i$ should not be confused with the subscripts $_i$.}
\bq\label{eq:example_c_2}
\phi_{\rm Lag}(\vk) = \phi_{\rm Eul}(\vk) + \int_{\vp} \frac{\vp(\vk-\vp)}{p^2} \delta_m^{(1)}(\vp) \phi_{\rm Eul}(\vk - \vp).
\eq
In these equations we have marked when $\phi$ is evaluated in Lagrangian or Eulerian space; in the following, $\phi$ is always the Eulerian one and we drop the subscripts to ease the notation. The corresponding term $\propto b_1^2 b_\phi \fnl$ in Eq.~(\ref{eq:bggg_model_NG}) is then given by two contributions
\bq\label{eq:example_c_3}
&& b_1^2 b_\phi \fnl \Bigg[\int_{\vp}\frac{\vp(\vk-\vp)}{p^2}\big<\delta_m(\vk_a)\delta_m(\vk_b)\delta_m^{(1)}(\vp)\phi(\vk_c-\vp)\big> + {\rm (2\ perm.)}\Bigg] \nonumber \\
&+& b_1^2 b_\phi \fnl \Bigg[\frac{\big<\delta_m(\vk_a)\delta_m(\vk_b)\delta_m^{(1)}(\vk_c)\big>}{\mathcal{M}_c} + {\rm (2\ perm.)}\Bigg]. \nonumber \\
\eq
The first of these can be worked out analogously to as in Eq.~(\ref{eq:example_a_2}). The derivation of the second contribution is in all analogous to the derivation of the tree-level matter bispectrum in perturbation theory \cite{Bernardeau/etal:2002}, just with the caveat that one of the modes in the three-point function is always a linear one, i.e., only two modes are expanded as $\delta_m(\vk) = \delta_m^{(1)}(\vk) + \delta_m^{(2)}(\vk) + \cdots$, with $\delta_m^{(2)}(\vk) = \int_{\vr} F_2(\vr, \vk - \vr)\delta_m^{(1)}(\vr)\delta_m^{(1)}(\vk-\vr)$. From this point onwards, the derivation follows straightforwardly and yields the corresponding term in Eq.~(\ref{eq:bggg_model_NG}) (always in keeping with the assumption of sufficiently narrow bins in Eq.~(\ref{eq:bggg_estimator}) to justify skipping performing the bin averages explicitly).

\section{The impact of covariance composition}
\label{app:cov}

In this appendix, we outline the derivation of the covariance matrix of the combined power spectrum and bispectrum data vector. We also show the impact that varying levels of completion of the covariance calculation have on the resulting $\fnl$ bounds. 

The covariance of the galaxy auto-power spectrum is defined as 
\bq\label{eq:covPP_derivation_1}
\cov^{P^{\rm AA}P^{\rm AA}}(k_1, k_2) &=& \Big<\hat{P}_{gg}^{\rm AA}(k_1)\hat{P}_{gg}^{\rm AA}(k_2)\Big> - P_{gg}^{\rm AA}(k_1)P_{gg}^{\rm AA}(k_2) \nonumber \\
&=& \frac{1}{V_s^2V_{k_1}V_{k_2}} \int_{k_1} {\rm d}^3\vk_a \int_{k_2} {\rm d}^3\vk_b \Big<\delta_g^{\rm A}(\vk_a)\delta_g^{\rm A}(-\vk_a)\delta_g^{\rm A}(\vk_b)\delta_g^{\rm A}(-\vk_b)\Big> - P_{gg}^{\rm AA}(k_1)P_{gg}^{\rm AA}(k_2) \nonumber \\
&=& \frac{1}{V_s^2V_{k_1}V_{k_2}} \int_{k_1} {\rm d}^3\vk_a \int_{k_2} {\rm d}^3\vk_b (2\pi)^6[P_{gg}^{\rm AA}(k_a)]^2 \Big[\delta_D(\vk_{ab})\delta_D(-\vk_{ab}) + \delta_D(\vk_a - \vk_b)\delta_D(\vk_b - \vk_a)\Big]\nonumber \\
&=& \frac{2 (2\pi)^3}{V_sV_{k_1}V_{k_2}} \delta_{k_1k_2}\int_{k_1} {\rm d}^3\vk_a  [P_{gg}^{\rm AA}(k_a)]^2 \nonumber \\
&=& \frac{2 (2\pi)^3}{V_sV_{k_1}} \delta_{k_1k_2} [P_{gg}^{\rm AA}(k_1)]^2,
\eq
where in the third equality we have used Wick's theorem and discarded the contribution from the connected four-point function, which is negligible on the large scales where the $\fnl$ contribution is important; in the fourth equality we have integrated using the Dirac delta functions, which imposes the constrain $\delta_{k_1k_2}$ that the two wavenumbers must belong to the same bin for the result to be non-zero (we have used also that $\delta_D(0) \equiv V_s/(2\pi)^3$); in the last equality we have assumed sufficiently narrow bins to take the power spectrum out of the integral. The steps above are for the derivation of the covariance of the auto-power spectrum of galaxy sample A, which gives the upper left entry in Eq.~(\ref{eq:covPP}); the remainder of the entries of the multitracer power spectrum covariance follow analogously. 

The covariance of the galaxy bispectrum is defined as (recall, in the main body of the paper, we considered the bispectrum of the subsample A)
\bq\label{eq:covBB_derivation_1}
&& \cov^{B^{\rm AAA}B^{\rm AAA}}(k_1, k_2, k_3,k_1', k_2', k_3') = \Big<\hat{B}^{\rm AAA}_{ggg}(k_1, k_2, k_3)\hat{B}^{\rm AAA}_{ggg}(k_1', k_2', k_3')\Big> \nonumber \\
&-& {B}^{\rm AAA}_{ggg}(k_1, k_2, k_3){B}^{\rm AAA}_{ggg}(k_1', k_2', k_3') \nonumber \\
&=& \frac{1}{V_s^2V_{123}V_{1'2'3'}} \int_{k_1} {\rm d}^3\vk_a \int_{k_2} {\rm d}^3\vk_b \int_{k_3} {\rm d}^3\vk_c \int_{k_1'} {\rm d}^3\vk_a' \int_{k_2'} {\rm d}^3\vk_b' \int_{k_3'} {\rm d}^3\vk_c' \nonumber \\
&\times& \Big<\delta^{\rm A}_g(\vk_a)\delta^{\rm A}_g(\vk_b)\delta^{\rm A}_g(\vk_c)\delta^{\rm A}_g(\vk_a')\delta^{\rm A}_g(\vk_b')\delta^{\rm A}_g(\vk_c')\Big> \delta_D(\vk_{abc}) \delta_D(\vk_{abc}')  - {B}^{\rm AAA}_{ggg}(k_1, k_2, k_3){B}^{\rm AAA}_{ggg}(k_1', k_2', k_3'). \nonumber \\
\eq
Using Wick's theorem, the six-point function contribution can be split into terms proportional to the product of three two-point functions, called $PPP$ term; terms proportional to the product of two three-point functions, called the $BB$ term; terms proportional to the product of a two-point function and a four-point function, called the $TP$ term; and finally, the connected six-point function term. In the main body of the paper, we considered only the contribution from the $PPP$ and the $BB$ term, which are the most straightforward to evaluate and yield Eqs.~(\ref{eq:covBBPPP}) and (\ref{eq:covBBBB}), respectively (see e.g. Refs.~\cite{2006PhRvD..74b3522S, 2017PhRvD..96b3528C, sqbcov} for detailed derivations/discussions). As argued recently in Refs.~\cite{2017PhRvD..96b3528C, sqbcov}, the contribution from the connected six-point function is expected to be negligible on the large scales and squeezed configurations relevant for $\fnl$ constraints. The $TP$ term has the same structure of ratios $U / V_{123}V_{1'2'3'}$ and the same powers of power spectra as the $BB$ term, so they display normally the same order of magnitude (especially in the squeezed limit \cite{sqbcov}). Below, we measure roughly the importance of the $TP$ term by doubling the contribution from the $BB$ term. 

Finally, the cross-covariance of the power spectrum and matter bispectrum is defined as 
\bq\label{eq:covBP_derivation_1}
&& \cov^{B^{\rm AAA}P^{\rm AB}}(k_1, k_2, k_3,k_1') = \Big<\hat{B}^{\rm AAA}_{ggg}(k_1, k_2, k_3)\hat{P}^{\rm AB}_{gg}(k_1')\Big> - {B}^{\rm AAA}_{ggg}(k_1, k_2, k_3){P}^{\rm AB}_{gg}(k_1') \nonumber \\
&=& \frac{1}{V_s^2V_{123}V_{k_1'}} \int_{k_1} {\rm d}^3\vk_a \int_{k_2} {\rm d}^3\vk_b \int_{k_3} {\rm d}^3\vk_c \int_{k_1'} {\rm d}^3\vk_a' \Big<\delta_g^{\rm A}(\vk_a)\delta_g^{\rm A}(\vk_b)\delta_g^{\rm A}(\vk_c)\delta_g^{\rm A}(\vk_a')\delta_g^{\rm B}(-\vk_a')\Big> \delta_D(\vk_{abc}) \nonumber \\
&& - {B}^{\rm AAA}_{ggg}(k_1, k_2, k_3){P}^{\rm AB}_{gg}(k_1') \nonumber \\
&=& \frac{(2\pi)^3}{V_sV_{k_1'}} \Bigg[ \delta_{k_1k_1'} \bigg(P_{gg}^{\rm AA}(k_1) B_{ggg}^{\rm AAB}(k_2,k_3,k_1) + P_{gg}^{\rm AB}(k_1) B_{ggg}^{\rm AAA}(k_2,k_3,k_1) \bigg) \nonumber \\
&&\ \ \ \ \ \ \ + \delta_{k_2k_1'} \bigg(P_{gg}^{\rm AA}(k_2) B_{ggg}^{\rm AAB}(k_1,k_3,k_2) + P_{gg}^{\rm AB}(k_2) B_{ggg}^{\rm AAA}(k_1,k_3,k_2) \bigg) \nonumber \\
&&\ \ \ \ \ \ \ + \delta_{k_3k_1'} \bigg(P_{gg}^{\rm AA}(k_3) B_{ggg}^{\rm AAB}(k_1,k_2,k_3) + P_{gg}^{\rm AB}(k_3) B_{ggg}^{\rm AAA}(k_1,k_2,k_3) \bigg) \Bigg],
\eq
where the third equality follows from Wick's theorem (dropping the contribution from the connected five-point function) plus derivation steps analogous to those already sketched above. This expression, as written, corresponds to the cross-covariance of the bispectrum with the $\hat{P}^{AB}$ part of the data vector, but the steps for the $\hat{P}^{AA}$ and $\hat{P}^{BB}$ are analogous. Further, $B_{ggg}^{\rm AAB}$ denotes a cross-bispectrum defined as $\big<\delta_g^{\rm A}(\vk_a)\delta_g^{\rm A}(\vk_b)\delta_g^{\rm B}(\vk_c)\big> = (2\pi)^3B_{ggg}^{\rm AAB}(\vk_a, \vk_b, \vk_c)\delta_D(\vk_{abc})$.

It is important to mention also that we do not account for any super-sample covariance (SSC) contributions that arise from modes with wavelengths larger than the size of the surveys \cite{takada/hu:2013, li/hu/takada, 2018JCAP...02..022L, completessc, 2016arXiv161205958L, 2020arXiv200514677C} (see also Refs.~\cite{2006MNRAS.371.1188H, 2006PhRvD..74b3522S, 2009MNRAS.395.2065T, 2012JCAP...04..019D, 2007NJPh....9..446T, 2013MNRAS.429..344K, 2009ApJ...701..945S, 2003ApJ...584..702H}). This is justified for the power spectrum part, for which on the large scales of interest, the SSC contributions are negligible compared to the dominant $\cov^{PP}$. Similarly, for the bispectrum covariance, Refs.~\cite{2018PhRvD..97d3532C, sqbcov} conclude that SSC contributions are only a negligible contribution compared to the $PPP$, $BB$ and $TP$ terms. The SSC contributions to the cross-covariance $\cov^{BP}$ have not been studied with as much detail, although Ref.~\cite{2018PhRvD..97d3532C} finds that it could be more important than for $\cov^{BB}$ for the matter bispectrum; Ref.~\cite{sqbcov} further suggests that although it can have a small impact on signal-to-noise ratios, its presence may help regularize the covariance matrix and keep it positive-definite (see below). 

\begin{figure}[t!]
	\centering
	\includegraphics[width=9cm]{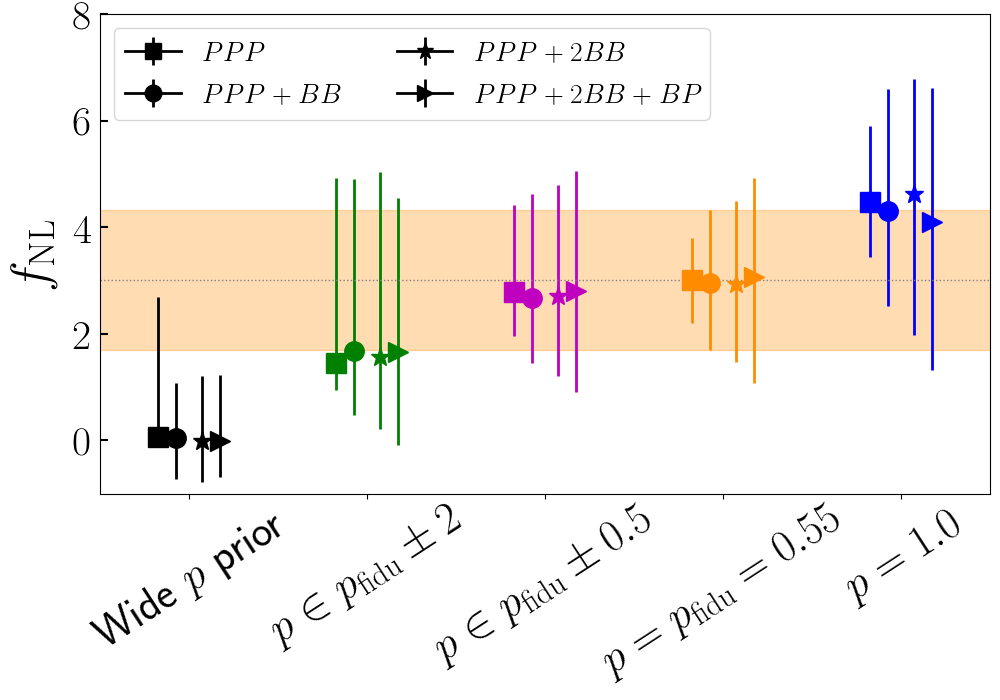}
	\caption{Marginalized $1\sigma$ constraints on $\fnl$ from combined $P_{gg} + B_{ggg}$ analysis with parametrization 1, $b_\phi = 2\delta_c(b_1-p)$. The different colors correspond to  the different priors on $p$ indicated in the x-axis. The different symbols show the constraints obtained with varying levels of composition of the covariance matrix, as labeled (see text for more details). The $PPP+BB$ is the default case used in the main body of the paper. The orange band marks the constraints with $PPP+BB$ and $p = p_{\rm fidu}=0.55$.}
\label{fig:sigmas_cov}
\end{figure}

Figure \ref{fig:sigmas_cov} helps understand the relative importance of the various contributions to the covariance matrix. The figure shows the marginalized $1\sigma$ constraints on $\fnl$ obtained from combined $P_{gg} + B_{ggg}$ analyses with parametrization 1, $b_\phi = 2\delta_c(b_1-p)$. The different symbols show the result for varying levels of composition of the covariance matrix. The one labeled $PPP+BB$ is our default calculation with $\cov^{BB} = \cov^{BB}_{PPP} + \cov^{BB}_{BB}$ and $\cov^{BP}=0$. The one labeled $PPP$ drops the $BB$ term from the bispectrum covariance, which as expected, results in an overestimate of the constraining power of the data, i.e., smaller uncertainties $\sigma_{\fnl}$. For example, for the $p = p_{\rm fidu} = 0.55$ case, dropping the $BB$ term shrinks $\sigma_{\fnl}$ by roughly $60\%$. Note that many studies in the literature include only the $PPP$ contribution, and may therefore represent too optimistic scenarios. Further, the impact of doubling the size of the $BB$ term (labeled $PPP + 2BB$), results in roughly $15\%$ increase in $\sigma_{\fnl}$ for the cases $p \in p_{\rm fidu} \pm 0.5$ and $p = p_{\rm fidu} = 0.55$; recall, this represents a very approximate way to account for the missing $TP$ term. Finally, further adding the power spectrum-bispectrum cross-covariance can increase $\sigma_{\fnl}$ by about $15\%$ to $30\%$ for the same two $p \in p_{\rm fidu} \pm 0.5$ and $p = p_{\rm fidu} = 0.55$ cases, respectively.

The result depicted in Fig.~\ref{fig:sigmas_cov} motivates future forecast studies on $\fnl$ to begin incorporating covariance contributions beyond the $PPP$ (as it is commonly done), since they may significantly degrade the constraining power on $\fnl$. On the other hand, robust forecasts on $\fnl$ should also include modelling of systematic errors, which add to the total error budget, and therefore reduce the importance of the statistical covariance matrix. A main message here is that future robust survey-specific forecasts on $\fnl$ should perform convergence tests similar to that depicted in Fig.~\ref{fig:sigmas_cov} to guarantee that the constraining power of the data is not being overestimated.

As a final remark, we would like to mention that, at least in the context of our idealized forecast setup, the covariance matrix ceases to be positive definite if the cross-covariance term is taken into account without doubling the contribution from the $BB$ term. We did not investigate this issue further, but speculate that it could be either due to the missing contribution from the connected five-point function (which could include SSC terms) in the cross-covariance part, and/or that the bispectrum covariance matrix must meet some level of completeness (i.e., contain the $TP$ terms) before the cross-covariance part is added without spoiling positive-definiteness. We have further found that this is an issue that develops only if $k_{\rm max} \gtrsim 0.14 h / \rm {Mpc}$, which gives some support to the explanation that this could be due to missing contributions that grow important on smaller scales.

\bibliographystyle{utphys}
\bibliography{REFS}

\end{document}